\documentclass[12pt]{emulateapj}

\slugcomment{Accepted, to appear in ApJ, v693, 2009 March.}
\pdfoutput=1
\usepackage{natbib}
\usepackage{epsfig}
\begin{document}

\defcitealias{2008arXiv0805.2950R}{Paper I}
\newcommand\etal{{\it et al.}}
\newcommand\eg{{\it e.g.,~}}
\newcommand\ie{{\it i.e.,~}}

\title{On the time variability of geometrically-thin black hole accretion disks II: Viscosity-induced global oscillation modes in simulated disks}

\author{Sean M. O'Neill, Christopher S. Reynolds, M. Coleman Miller}
\affil{Department of Astronomy and Maryland Center for Theory and Computation, \\ University of Maryland, College Park, MD 20742}

\begin{abstract}
We examine the evolution and influence of viscosity-induced diskoseismic modes in simulated black hole accretion disks.  
Understanding the origin and behavior of such oscillations will help us to evaluate their potential role in producing astronomically observed high-frequency quasi-periodic oscillations in accreting black hole binary systems.

Our simulated disks are geometrically-thin with a constant half-thickness of five percent the radius of the innermost stable circular orbit.  
A pseudo-Newtonian potential reproduces the relevant effects of general relativity, and an alpha-model viscosity achieves angular momentum transport and the coupling of orthogonal velocity components in an otherwise ideal hydrodynamic numerical treatment.

We find that our simulated viscous disks characteristically develop and maintain trapped global mode oscillations with properties similar to those expected of trapped g-modes and inner p-modes in a narrow range of frequencies just below the maximum radial epicyclic frequency.  
Although the modes are driven in the inner portion of the disk, they generate waves that propagate at the trapped mode frequencies out to larger disk radii.  
This finding is contrasted with the results of global magnetohydrodynamic disk simulations, in which such oscillations are not easily identified.  
Such examples underscore fundamental physical differences between accretion systems driven by the magneto-rotational instability and those for which alpha viscosity serves as a proxy for the physical processes that drive accretion, and we explore potential approaches to the search for diskoseismic modes in full magnetohydrodynamic disks.
\end{abstract}


\section{Introduction}\label{sec:introduction}

Since the detection of the first high-frequency quasi-periodic oscillations (HFQPOs) from black hole candidate GRS 1915+105 \citep{1997ApJ...482..993M}, much effort has been made to relate such oscillations to natural accretion disk frequencies.
Some early analysis by \citet{1997ApJ...477L..91N} suggested that these HFQPOs were manifestations of global oscillation modes in galactic black hole binaries (GBHBs), the theory of which had been explored extensively by, for example, \citet{1987PASJ...39..457O} and \citet{1991ApJ...378..656N,1992ApJ...393..697N,1993ApJ...418..187N}.
The discovery by \citet{2001ApJ...552L..49S} of a pair of HFQPOs in GRO J1655-40 with an approximate 3:2 frequency ratio, however, lent support to an alternative parametric resonance model \citep{2001A&A...374L..19A} in which HFQPOs result from resonance between orbital and radial epicyclic motion of disk material.
This and similar resonance models have the advantage of naturally generating the small-integer frequency ratios seen in GRO J1655-40 and some subsequently observed sources \citep[for a summary of current HFQPO observations, see][]{2006ARA&A..44...49R}.
Still, parametric resonance models have yet to incorporate convincing physical mechanisms by which to excite HFQPOs \citep{2008NewAR..51..855R}, and it has been noted by \citet{2008GApFD.102...75O} that multiple global oscillation modes of differing mode number produce a 3:2 frequency ratio equally well. 
A detailed physical understanding of such oscillations is crucial since their observed frequencies ($\sim 100$ Hz) are comparable to orbital frequencies near the innermost stable circular orbit (ISCO) of stellar-mass black holes.
Interpreted correctly, HFQPOs thus have the potential to tell us much about the inner portions of accretion disks and the black holes they orbit.

While numerical simulations have great promise to elucidate the nature of disk oscillations, they unfortunately have failed thus far to produce convincing, identifiable HFQPOs.
Despite some preliminary claims of HFQPO generation in relatively low-resolution simulations by \citet{2004PASJ...56..931K}, subsequent numerical magnetohydrodynamic (MHD) studies have shown that simulated HFQPOs typically are transient \citep{2006ApJ...651.1031S}, require external driving \citep{2006astro.ph.11269C}, or, in the case of the first paper in this series, remain completely undetected \citep[hereafter \citetalias{2008arXiv0805.2950R}]{2008arXiv0805.2950R}.
Interestingly, the simulations of ideal {\it hydrodynamic} disks in \citetalias{2008arXiv0805.2950R} did generate trapped gravity-driven (\ie g-mode) global disk oscillations such as those described in \citet{1992ApJ...393..697N}, but these oscillations were not seen in their otherwise comparable MHD disks.
In fact, oscillations of the amplitude seen in X-ray observations or in their hydrodynamic disks would have fallen below the level of turbulent noise generated in the MHD case, so they could not determine whether the modes were hidden or actively damped by turbulence in the manner discussed by \citet{2006ApJ...645L..65A}.
Regardless, it is clear that the prototypical MHD disks of \citetalias{2008arXiv0805.2950R} failed to excite to a detectable level either the global diskoseismic oscillations of \citet{1993ApJ...418..187N} or the parametric resonance instability of \citet{2001A&A...374L..19A}.

In this paper, we describe complementary work to \citetalias{2008arXiv0805.2950R} in the form of simulations with oscillations induced by the viscous tapping of orbital energy.
While the standard physical model for black hole accretion is predicated upon the magneto-rotational instability \citep[MRI,][]{1991ApJ...376..214B}, an intrinsically MHD process that naturally generates turbulence and the transport of angular momentum, the traditional alternative to full MHD simulations has been to mimic the influence of magnetic fields through the introduction of an ``alpha-model'' viscosity \citep{1973A&A....24..337S}.
This approach subsumes all physical details of accretion into a single dimensionless viscous parameter $\alpha$ designed to achieve the appropriate global level of angular momentum transport.
This approach is by no means completely equivalent to a full MHD treatment, as \citet{1998RvMP...70....1B} and \citet{2008MNRAS.383..683P} note, and there even remain some basic order-of-magnitude discrepancies between values of $\alpha$ inferred from observation and those derived from MHD simulations \citep{2007MNRAS.376.1740K}.
Still, studying alpha-disks provides us with a method by which to evaluate the influence of viscosity independent of MRI-driven turbulence and other typically complex behaviors of fully MHD disks.

\citet{2000ApJ...537..922O} have provided a linear analysis of normal modes in viscous, rotating, Newtonian fluids that is applicable to our simulated viscous accretion disks.
In particular, they show that the presence of viscosity should cause the fundamental g-modes in rotating disks to grow at a rate that scales with the characteristic orbital frequency in the system.
In relativistic or pseudo-Newtonian gravitational potentials, these g-modes are predicted to be non-evanescent at radii where $|\omega| < \kappa$, where $\omega$ is the wave frequency, and $\kappa$ is the radial epicyclic frequency [see \citet{1991ApJ...378..656N,1992ApJ...393..697N,2000ApJ...537..922O}, or Section 2.2 of \citetalias{2008arXiv0805.2950R} for the appropriate dispersion relations and derivations].
In black hole accretion disks, this means that the modes are trapped just under the maximum radial epicyclic frequency $\kappa_{\rm max}$.
Similarly, \citet{2000ApJ...537..922O} find that viscosity also causes the inner pressure-driven (p-mode) oscillations to grow for $\kappa_{\rm max} > |\omega| > \kappa$, although the successful trapping of such modes depends strongly upon on the nature of the inner boundary of the disk.

While the viscosity-induced trapped g-mode of \citet{2000ApJ...537..922O} has never been identified explicitly in simulations, numerical models of viscous hydrodynamic disks have generated identifiable waves at frequencies comparable to this mode.
In an early numerical analysis of axisymmetric, vertically integrated (\ie 1D) disks, \citet{1992PASJ...44..529H} found that viscosity above a critical value $\alpha \sim 0.1$ caused global disk oscillations near $\kappa_{\rm max}$.
Likewise, \citet{1995ApJ...441..354C} and \citet{1996MNRAS.283..919M} identified global oscillations near $\kappa_{\rm max}$ in vertically integrated disk simulations for a range of moderate accretion rates.
This work was followed by the 2D simulations of \citet{1997MNRAS.286..358M}, which focused on convection in optically thick disks but also found oscillations near $\kappa_{\rm max}$, particularly for low accretion rates and large viscosities.
More recently, \citet{2008arXiv0805.0598M} revisited the vertically integrated models of \citet{1996MNRAS.283..919M}, pointing out that waves propagating from the inner portions of the disk could easily be locally super-Keplerian. 
All of these studies associate the observed signals with radial inertial-acoustic oscillations corresponding to the previously mentioned inner p-modes.
This is a particularly valid interpretation in the case of vertically integrated disks where motion is constrained to the radial dimension, but distinguishing between trapped g- and inner p-modes in 2D viscous disks is not as straightforward, as we shall discuss in this work.

Our simulations of viscous accretion disks are intended to complement this previous work by exploring in more detail how viscosity can induce diskoseismic modes in accretion disks and how these modes affect the body of the disk.
First, we seek to discover whether we can produce and identify in our models any of the viscosity-induced modes of \citet{2000ApJ...537..922O}.
In particular, we are interested in the trapped global g-modes since they exist in a narrow frequency range near $\kappa_{\rm max}$, the value of which in principle can be used as a diagnostic of the fundamental physical properties of the black hole.
Since these g-modes are trapped well away from the inner boundary of the disk, we also expect them to be less susceptible than inner p-modes to leakage across the ISCO.
We further examine how such modes generate waves that propagate through the entire body of the disk, far beyond the formal mode trapping region.
Since trapped g-modes were identified in the hydrodynamic simulations -- but not the MHD disks -- of \citetalias{2008arXiv0805.2950R}, we further seek to understand and evaluate the observed differences between viscous alpha-disks and full MHD models.

In $\S$ \ref{sec:modeling}, we outline the computational framework used and describe our simulated disk models.
In $\S$ \ref{sec:results}, we present the results of our simulations and discuss our identification of trapped diskoseismic modes and the effects of these modes on viscous disks.
We place our findings in a broader context in $\S$ \ref{sec:discussion}, comparing our results to previous work, and present our conclusions in $\S$ \ref{sec:conclusions}.

\section{Modeling Viscous Disks}\label{sec:modeling}
\subsection{Numerical Methods}\label{sec:numerical}
To simulate the evolution of viscous accreting systems, we have adapted the ZEUS-MP code (version 2), the basic workings of which are described in \citet{1992ApJS...80..753S,1992ApJS...80..791S}, \citet{1992ApJS...80..819S}, and more recently in \citet{2006ApJS..165..188H}.
This code employs an Eulerian finite difference scheme to solve to second-order accuracy the equations of ideal compressible fluid dynamics.
For our purposes, we run ZEUS-MP in pure hydrodynamic mode using cylindrical coordinates ($r,z,\phi$).
The calculation is ``2.5 dimensional'', meaning that it enforces complete azimuthal symmetry while allowing a non-zero azimuthal velocity.
Our simulations feature a gamma-law gas equation of state ($p \propto \rho^\gamma$) with a constant $\gamma = 5/3$.
Zero-gradient outflow boundary conditions are enforced at each timestep in both the $r$ and $z$ directions.
Additionally, we employ a protection routine to impose a density floor $\rho_{\rm min}$ at a value $10^{-7}$ times that of the initial disk midplane density.

We have modified ZEUS-MP to incorporate additional physics relevant to the simulation of accretion disks.
While ZEUS-MP allows the inclusion of point-mass gravity, we have adjusted this to reflect a pseudo-Newtonian gravitational potential, such as that developed by \citet{1980A&A....88...23P}.
In this potential,
\begin{equation}
\label{eq:pnpot}
\Phi=-\frac{GM}{R-2r_{\rm g}}, \qquad r_{\rm g} \equiv \frac{GM}{c^2},
\end{equation}
where $R=\sqrt{r^2 + z^2}$ is the spherical radius.
This approach accurately reproduces the positions of the innermost stable circular orbit (ISCO) at $r=6r_{\rm g}$ and marginally bound orbit at $r=4r_{\rm g}$ for a Schwarzschild black hole.
Additionally, we have added to ZEUS-MP the aforementioned ``$\alpha$-model'' \citep{1973A&A....24..337S} prescription for viscosity in what is otherwise an ideal hydrodynamic system.
This modification consists of introducing a kinematic viscosity of the form
\begin{equation}
\label{eq:alphavisc}
\nu = \alpha c_{\rm s} H
\end{equation}
where
$\alpha$ is a dimensionless constant, $c_{\rm s}$ is the local sound speed, and $H \sim c_{\rm s}r/v_{\rm \phi}$ is the scale height of the disk, which we can express in terms of the sound speed, cylindrical radius, and azimuthal disk velocity ($v_{\rm \phi}$).
This model viscosity is applied as a correction to the force update in ZEUS-MP and directly updates the velocity components exclusively according to
\begin{equation}
\frac{\partial (\rho {\vec v})}{\partial t} = \mathbf{\nabla} \cdot \mathbf{\sigma} ,
\end{equation}
where the components of the viscous stress tensor $\mathbf{\sigma}$ are given in \citet{1959flme.book.....L}, for example.
Since we include viscosity only as a means to couple velocity components and to transport angular momentum, we assume that the dissipated heat is radiated away instantaneously and thus remove it from the system.
For numerical stability, the viscosity update must take place over a timescale less than or equal to the viscous timestep, given by
\begin{equation}
\label{eq:visctimestep}
\Delta t_{\rm visc} = C_{\rm visc} \min \left( \frac{\Delta x_{\rm i}^2}{\nu}\right),
\end{equation} 
where $C_{\rm visc}$ is a stability constant ($C_{\rm visc} \sim 0.1$) and $\Delta x_{\rm i}$ is the computational zone length in the $i$th dimension.
In practice, this criterion is met by subcycling the force update at a timestep 
\begin{equation}
\label{eq:subtimestep}
\Delta t_{\rm sub} = \frac{\Delta t_{\rm Cour}}{N} \le \Delta t_{\rm visc}, 
\end{equation}
where $\Delta t_{\rm Cour}$ is the standard Courant-Friedrichs-Levy timestep \citep[described in the context of ZEUS-MP by][]{2006ApJS..165..188H} and $N$ is the smallest positive integer to satisfy this condition.

\subsection{Simulated Disk Parameters}\label{sec:simulated}

The basic initialization template for our simulated accretion disks is identical to that of the 2D hydrodynamic disks discussed in \citetalias{2008arXiv0805.2950R}.
Additionally, we conduct two ``test'' simulations (described at the end of this section) to confirm that our results do not depend in an unphysical way upon the details of the initial conditions and computational grid size.
Since we do not simulate scale-dependent processes such as radiative cooling, our disks are fundamentally scale-free and we discuss them in natural units.

Our simulated disk density and pressure profiles are given by

\begin{equation}
\label{eq:dprofile}
\rho(r,z) = \rho_{\rm 0} \exp \left( -\frac{z^2}{2h_{\rm 1}^2}\right),
\end{equation}
and
\begin{equation}
\label{eq:pprofile}
p(r,z)=\frac{GMh_{\rm 2}^2}{(R-2r_{\rm g})^2R}~\rho(r,z) ,
\end{equation}
where $r$ is the cylindrical radius, $z$ is the vertical height above the disk midplane, and $R=\sqrt{r^2+z^2}$ is again the spherical radius.
The initial midplane value $\rho_{\rm 0}$ is independent of radius, as are the scale heights $h_{\rm 1}$ and $h_{\rm 2}$.
The disk is geometrically thin with a value of $h_{\rm 2}=0.3 r_{\rm g}$, leading to a ratio of $h_{\rm 2}/r_{\rm ISCO} = 0.05$ at the ISCO.
We set $h_{\rm 1} = 1.2h_{\rm 2}$ so that the disk is $\sim 20 \%$ too cold to maintain vertical hydrostatic equilibrium.
As a result of this setup, the initial disk collapses and oscillates before relaxing into an approximate steady state.
The initial velocity profile is entirely azimuthal with
\begin{equation}
v_{\rm \phi} = \frac{\sqrt{GMr}}{r-2r_{\rm g}} \qquad v_{\rm r} = v_{\rm z} =0,
\end{equation}
for $r \ge r_{\rm ISCO}$.
This corresponds to pure Keplerian motion in the disk midplane.

In all simulations, the computational grid spans a radial range of $r \in (4r_{\rm g},28r_{\rm g})$ and a vertical range of $z \in (-1.5r_{\rm g},1.5r_{\rm g})$.
The grid is populated by zones of uniform size $\Delta r \approx 2.3 \times 10^{-2}~r_{\rm g}$ and $\Delta z \approx 1.2 \times 10^{-2}~r_{\rm g}$, leading to a cell aspect ratio of 2:1.
This resolution provides $\sim 25$ vertical zones per pressure scale height $h_{\rm 2}$ and is thus sufficient to capture waves with wavelengths $\sim 4$ times smaller than the scale height.
The total duration of each simulation is $\sim 200 T_{\rm ISCO}$, where
\begin{equation}
\label{eq:tisco}
T_{\rm ISCO} \approx 61.6~GM/c^3
\end{equation}
is the orbital period at the ISCO.

The only input parameter adjusted across our models is the value of the dimensionless viscosity parameter $\alpha$.
As summarized in \citet{2007MNRAS.376.1740K}, evidence suggests that observed astrophysical accreting systems feature $\alpha \sim 0.1-0.4$.
Rather than restrict ourselves to this relatively narrow range of values, however, we instead examine an ensemble of simulated disks ranging from realistic values to completely inviscid disks.
Specifically, we choose model disks with $\alpha = \{0.1,~0.075,~0.05,~0.025,~0.01,~0.0\}$.
This enables us to explore how viscosity leads to the development and propagation of diskoseismic modes and how this behavior depends upon the strength of the viscosity.
Since the value of $\alpha$ is the single criterion by which we distinguish our disk models, we refer to them by this value prepended with an ``A'' (\eg model ``A0.05'' features $\alpha=0.05$).

Additionally, we briefly describe two ``test'' models designed to confirm that the behaviors of our disks are not unduly influenced by our specific simulation parameters.
In the model labeled ``EQ0.1'', we modify the basic disk template so that $h_{\rm 1}=h_{\rm 2}$ and move the inner edge of the disk to a distance $r \sim 1.33r_{\rm ISCO}$.
Since this disk is in vertical equilibrium and the inner edge of the disk is comfortably outside of the ISCO, this model helps us to gauge how our results depend upon the details of the initial disk perturbation.
Another model labeled ``GRD0.1'' features a grid that spans a radial range of $r \in (3.75r_{\rm g},28r_{\rm g})$ and a vertical range of $z \in (-1.52r_{\rm g},1.52r_{\rm g})$.
This model helps us to identify physically interesting disk behaviors and to isolate them from potentially unphysical oscillations caused by interactions with the inner radial computational grid boundary.
As their names suggest, both test models feature $\alpha=0.1$.

\section{Simulation Results}\label{sec:results}
\begin{figure}[t]
\centerline{
\includegraphics[type=pdf,ext=.pdf,read=.pdf,width=0.45\textwidth]{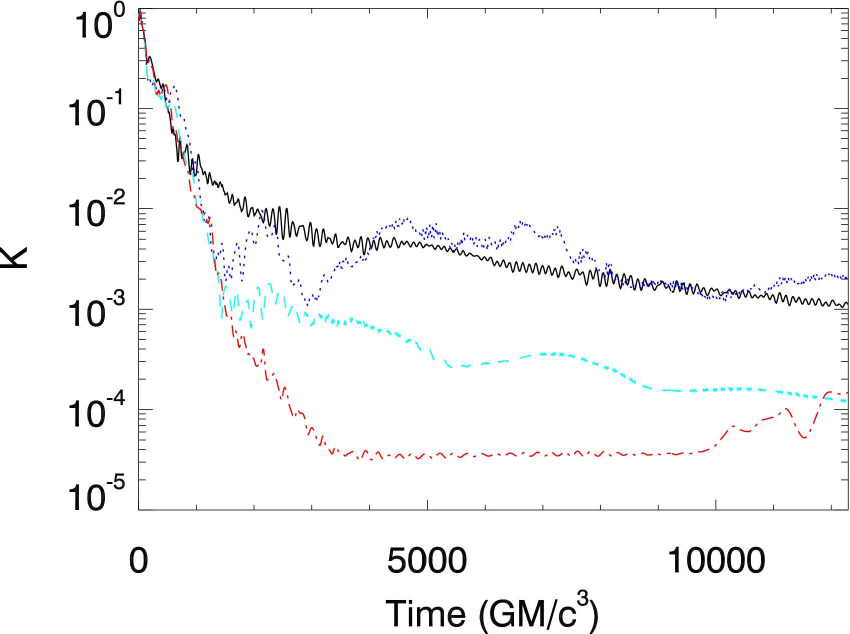}}
\caption[K Decay Rates]{Change of the quantity $K= \int_\mathcal{D} \rho v_{\rm z}^2 dV$ with time, normalized to its maximum value.  The integration domain $\mathcal{D}$ is the section $r \in (7r_{\rm g},14r_{\rm g})$.  Shown are models A0 (solid), A0.1 (dotted), A0.05 (dashed), and A0.01 (dot-dashed).  The long-term downward trend in A0 shows that the g-modes in this model are gradually losing energy that had been provided by the initial disk perturbation.  The viscous models, on the other hand, rapidly lose this initial energy.  The evolution of model A0.1, in particular, clearly shows that vertical energy is actively replenished in viscous disks.}
\label{fig:kdecayrates}
\end{figure}

We now discuss the evolution and analysis of our simulated disks.
While our models diverge as viscous effects become important, each disk is initially perturbed in the same way, and their early behaviors are quite similar.
As described in \citetalias{2008arXiv0805.2950R}, the initial disk setup is out of vertical equilibrium and so falls, rebounds, and eventually settles into an approximate steady state.
We evaluate the decay of the initial disk fluctuations by computing the quantity $K= \int_\mathcal{D} \rho v_{\rm z}^2 dV$, which is a measure of the energy in vertical disk oscillations.
Figure \ref{fig:kdecayrates} shows for models A0, A0.1, A0.05, and A0.01 the evolution of $K$ in time where the integration domain $\mathcal{D}$ covers the radial segment $r \in (7r_{\rm g},14r_{\rm g})$, away from the radial disk boundaries.
Although some of the intermediate viscosities are omitted from Figure \ref{fig:kdecayrates} to reduce visual clutter, the models shown bracket their behaviors.
We will discuss this figure in more detail in the following sections, but we note here that our different models feature very distinct evolutionary profiles in $K$.
This is hardly surprising since the viscous runs damp out some of the energy associated with the initial disk perturbation while also potentially introducing vertical oscillations through the mechanism of \citet{2000ApJ...537..922O}.
Rather than tailor our analysis of each individual model to that model's behavior in $K$, we conservatively restrict our discussion of all models to times after $t_{\rm relax} \sim 6 \times 10^3~GM/c^3$, corresponding to the approximate exponential decay time of $K$ in the inviscid model A0.
In \citetalias{2008arXiv0805.2950R}, this decay timescale was seen to increase with higher grid resolution, suggesting that numerical dissipation was responsible for damping out these initial oscillations.

\subsection{Inviscid Disks}
 
\begin{figure*}
\includegraphics[type=pdf,ext=.pdf,read=.pdf,width=0.54\textwidth]{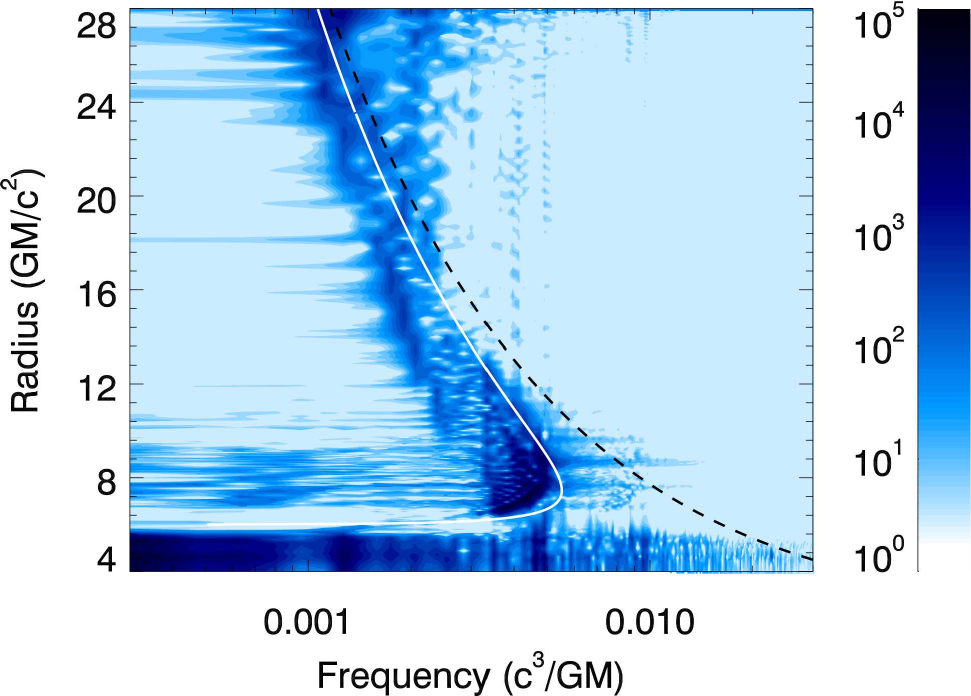}
\includegraphics[type=pdf,ext=.pdf,read=.pdf,width=0.54\textwidth]{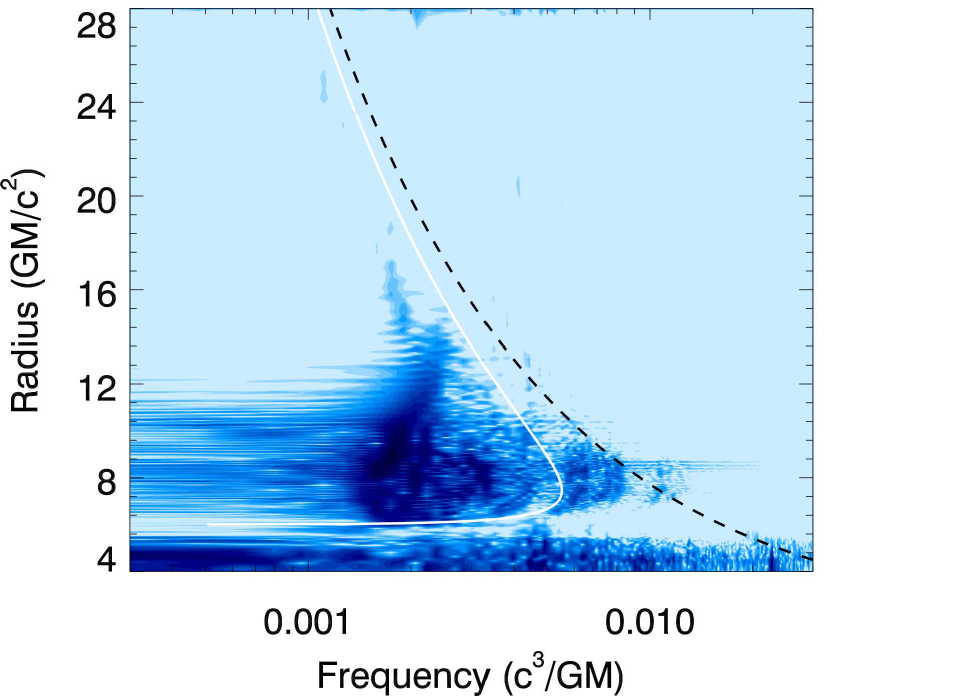}
\\[0.75em]
\includegraphics[type=pdf,ext=.pdf,read=.pdf,width=0.54\textwidth]{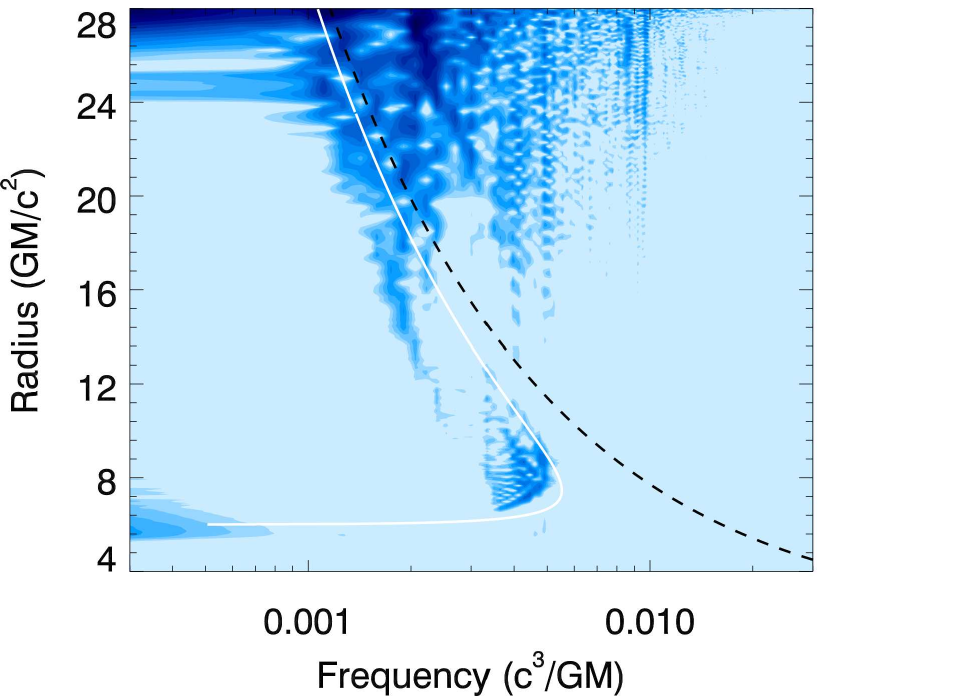}
\includegraphics[type=pdf,ext=.pdf,read=.pdf,width=0.54\textwidth]{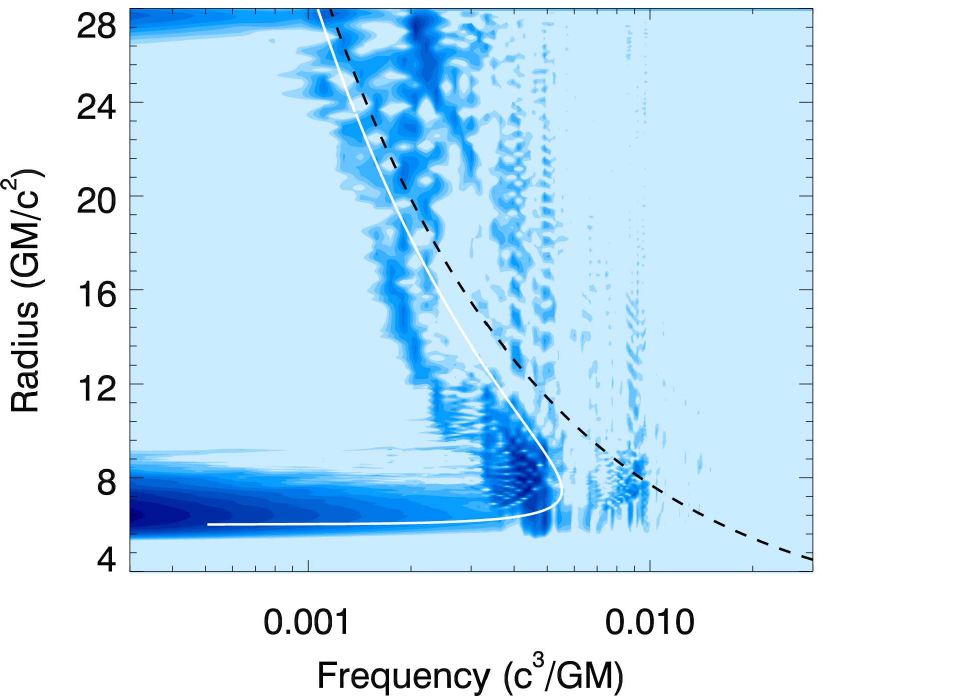}
\caption[Midplane PSDs for A0]{Midplane PSDs of radial velocity ({\it top-left}), vertical velocity ({\it top-right}), density ({\it bottom-left}) and pressure ({\it bottom-right}) for model A0.  Also shown are the radial epicyclic frequency (solid) and orbital frequency (dashed) for comparison.  The logarithmic colorbars are in arbitrary units and span five orders of magnitude.  The signal bounded on the right by the radial epicyclic frequency has the properties of a trapped g-mode, as described in \citetalias{2008arXiv0805.2950R}.}
\label{fig:a0psds}
\end{figure*}

Let us first review briefly the relevant characteristics of A0, the inviscid model similar to some of those described extensively in \citetalias{2008arXiv0805.2950R}.
Since we are most interested in physical processes that select specific frequencies, our primary method of analysis is to compute and examine the power spectral density (PSD), defined as $P(\nu) = \eta |\bar{f}(\nu)|^2$, where $\eta$ is a normalization constant and $\bar{f}(\nu)$ is the Fourier transform 
\begin{equation}
\bar{f}(\nu)=\int f(t)e^{-2\pi i \nu t} dt
\label{eq:ft}
\end{equation}
of a given time sequence $f(t)$.
Note that the PSDs in this paper are taken to be functions of the frequency $\nu$ instead of the angular frequency $\omega = 2\pi\nu$.
Figure \ref{fig:a0psds} shows the midplane (\ie $z=0$) PSDs of the radial and vertical velocities, pressure, and density in A0.
The absolute scales are arbitrary, but one can easily see in all four quantities an enhancement approximately bounded on the right by the radial epicyclic frequency.
Furthermore, the strongest enhancements lie just below the maximum radial epicyclic frequency, which, in the Paczynski-Wiita potential, is located at $\nu_{\rm max} = \omega_{\rm max}/2\pi \approx 5.5 \times 10^{-3}~{\rm c^3/GM}$ at a radius of $r_{\rm max} \approx 7.5~{\rm GM/c^2}$.
As \citetalias{2008arXiv0805.2950R} point out, these features have all of the expected characteristics of the trapped g-modes described by \citet{1992ApJ...393..697N}.
As noted, the evolution of $K$ in Figure \ref{fig:kdecayrates} illustrates that the energy in this trapped mode decays over time, having been introduced exclusively through the initial disk perturbation.

Additionally, we note some leakage of the g-mode signal both radially inward and outward from $r_{\rm max}$.
Most of the leakage that crosses the ISCO is expected to exit the grid, and the leakage radially outward from $r_{\rm max}$ is at least an order of magnitude less in strength than the trapped g-mode magnitude for all quantities.
We also note the presence of some broadband noise near the outermost disk radii.
This signal is caused by disk interactions with the outer grid boundary, as the resolution tests in \citetalias{2008arXiv0805.2950R} revealed, and similar features are seen in our simulations of viscous disks.

\subsection{Viscous Disks}

\begin{figure*}
\includegraphics[type=pdf,ext=.pdf,read=.pdf,width=0.54\textwidth]{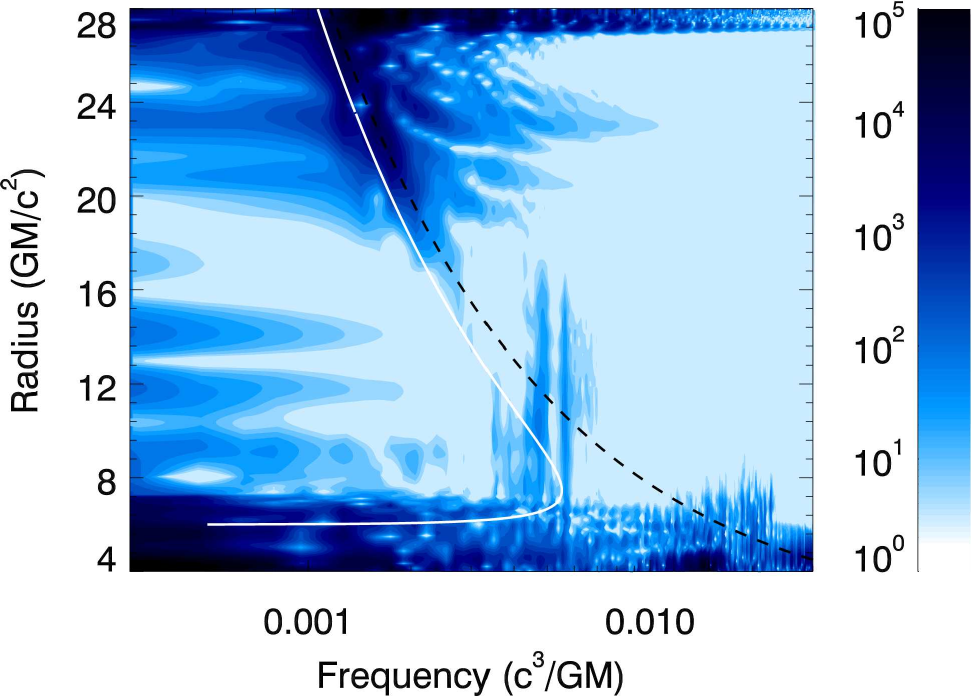}
\includegraphics[type=pdf,ext=.pdf,read=.pdf,width=0.54\textwidth]{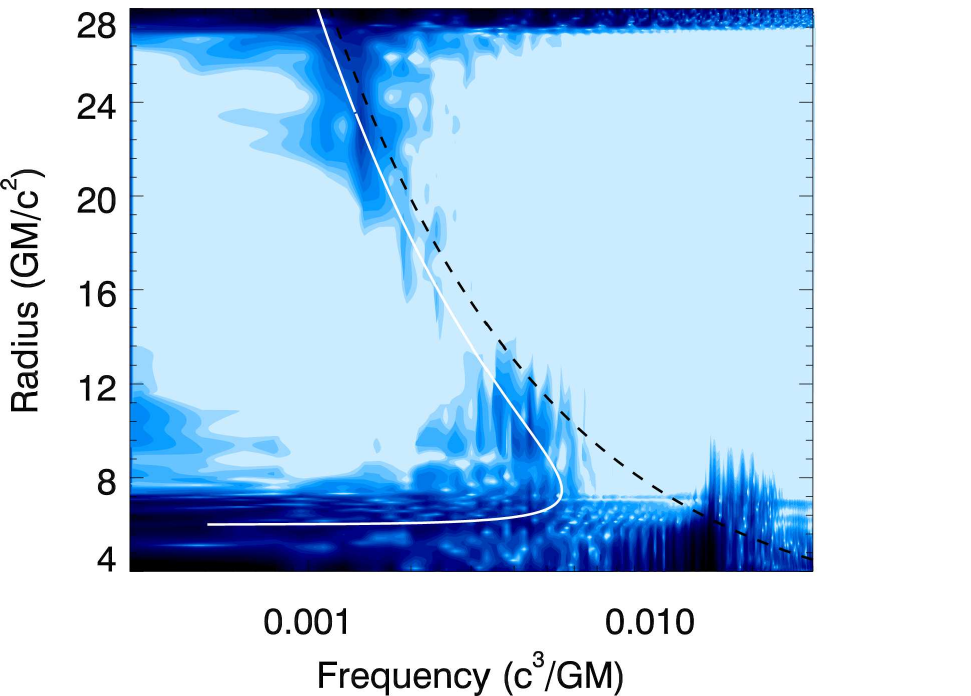}
\\[0.75em]
\includegraphics[type=pdf,ext=.pdf,read=.pdf,width=0.54\textwidth]{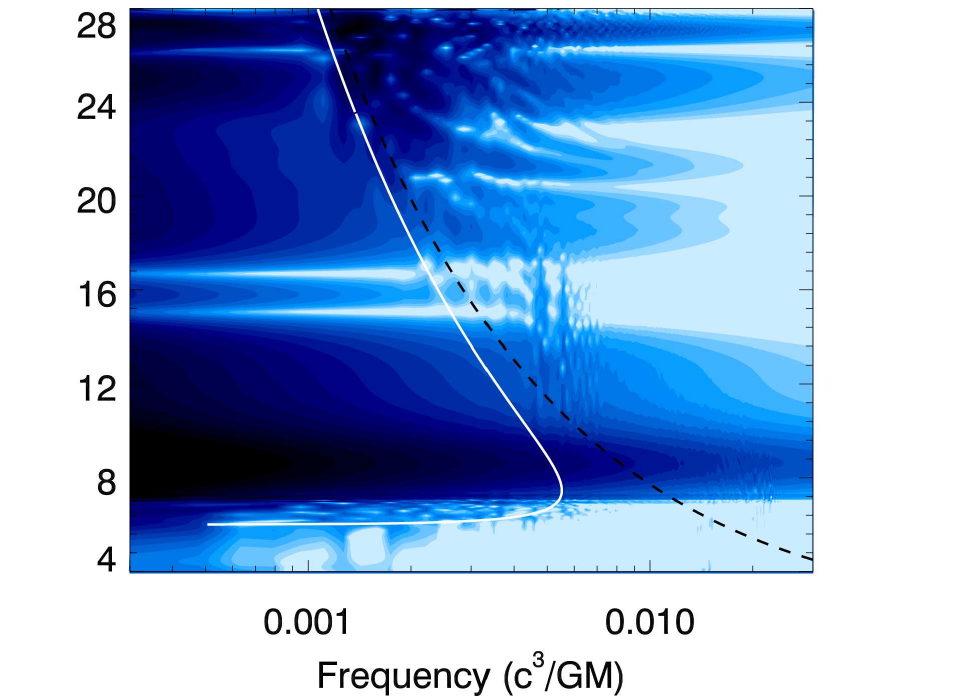}
\includegraphics[type=pdf,ext=.pdf,read=.pdf,width=0.54\textwidth]{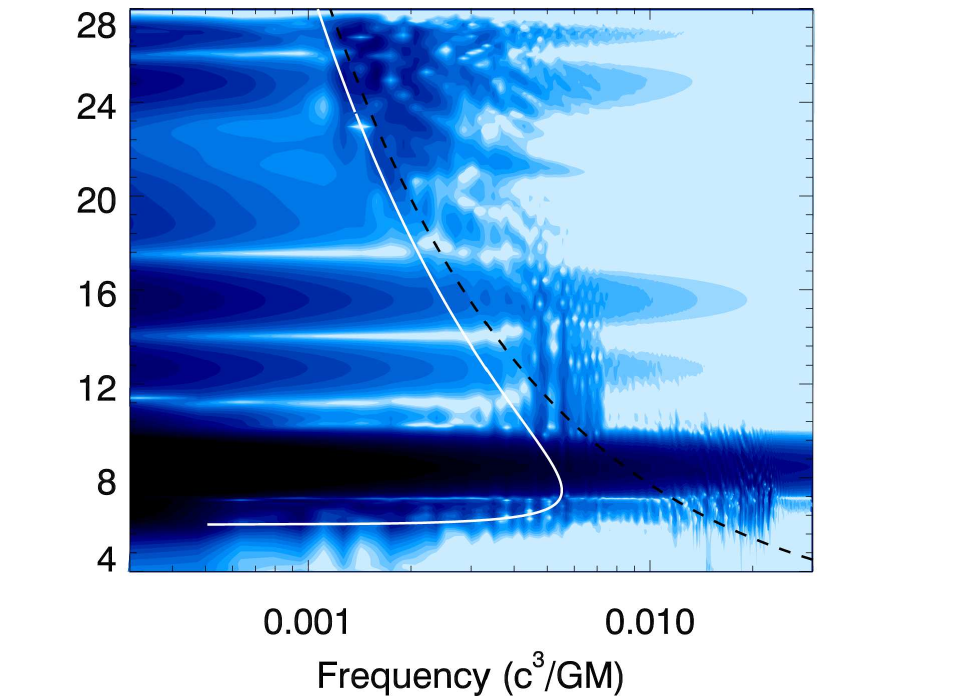}
\caption[Midplane PSDs for A0.1]{Midplane PSDs of radial velocity ({\it top-left}), vertical velocity ({\it top-right}), (decay-corrected) density ({\it bottom-left}) and (decay-corrected) pressure ({\it bottom-right}) for model A0.1.  Also shown are the radial epicyclic frequency (solid) and orbital frequency (dashed) for comparison.  The logarithmic colorbars span five orders of magnitude and are identical to those used in Figure \ref{fig:a0psds} to facilitate cross-comparison.  The set of vertical spikes just below the maximum radial epicyclic frequency indicate the presence of trapped modes that generate waves that pervade the disk.}
\label{fig:a1em1psds}
\end{figure*}

Considering now viscous disks, we will focus primarily on A0.1 for which the $\alpha$ value corresponds most closely to disk viscosities inferred from observations \citep{2007MNRAS.376.1740K}.
Figure \ref{fig:a1em1psds} shows the midplane PSDs for the radial and vertical velocities, pressure, and density in A0.1.
While the colorbars in both Figures \ref{fig:a0psds} and \ref{fig:a1em1psds} have arbitrary units, they are the same for both figures to facilitate cross-comparison of quantities between the two models.
First, we note that there is extensive broadband noise present, particularly for the density and pressure, in Figure \ref{fig:a1em1psds}.
That this noise is most pronounced at lower frequencies is consistent with the prediction by \citetalias{2008arXiv0805.2950R} that secular variation in the disk caused by the gradual loss of material through the radial outflow boundaries produces a signal that scales with $1/\omega^2$.
We have attempted to remove some of this noise from the density and pressure PSDs using a technique described in \citetalias{2008arXiv0805.2950R}.
In this approach, we divide these time series by an exponential decay function, choosing the time constant from a least-square fit to the data.
In A0.1, this secular trend amounts to a loss of only a few percent of the initial total disk mass during the period of analysis (\ie $t > t_{\rm relax}$), but Figure \ref{fig:a1em1psds} illustrates that a strong residual trend remains.
Fortunately, the velocity PSDs are affected by this variation only indirectly and, as such, require no secular correction.

The physically significant signal we see in A0.1 consists of a set of features located near $\nu \sim 5 \times 10^{-3}~{\rm c^3/GM}$ that extends radially (\ie vertically in Figure \ref{fig:a1em1psds}), occupying up to half of the disk radius.
Identifiable in all four quantities, this set of features is seen at a frequency range very near the maximum radial epicyclic frequency of the system at $\nu_{\rm max}$.
Although there is some variation in the signal profile for the four quantities shown in Figure \ref{fig:a1em1psds}, all signals characteristically feature one or two prominent broad ``spikes'' near $\nu_{\rm max}$ with multiple weaker peaks at adjacent lower frequencies.
The spikes in the velocity PSDs are seen down to the inner radius of the computational grid ($r=4~{\rm GM/c^2}$), while those for the pressure and density are difficult to identify inward of the broadband noise bands near $r \sim 7.2~{\rm GM/c^2}$.

\begin{figure*}
\includegraphics[type=pdf,ext=.pdf,read=.pdf,width=0.329\textwidth]{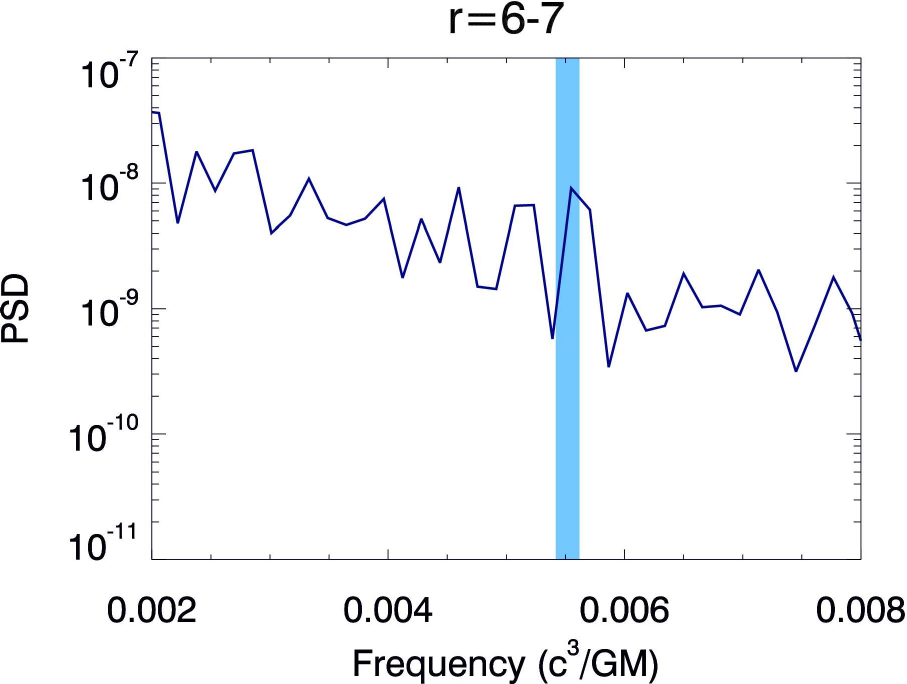}
\includegraphics[type=pdf,ext=.pdf,read=.pdf,width=0.329\textwidth]{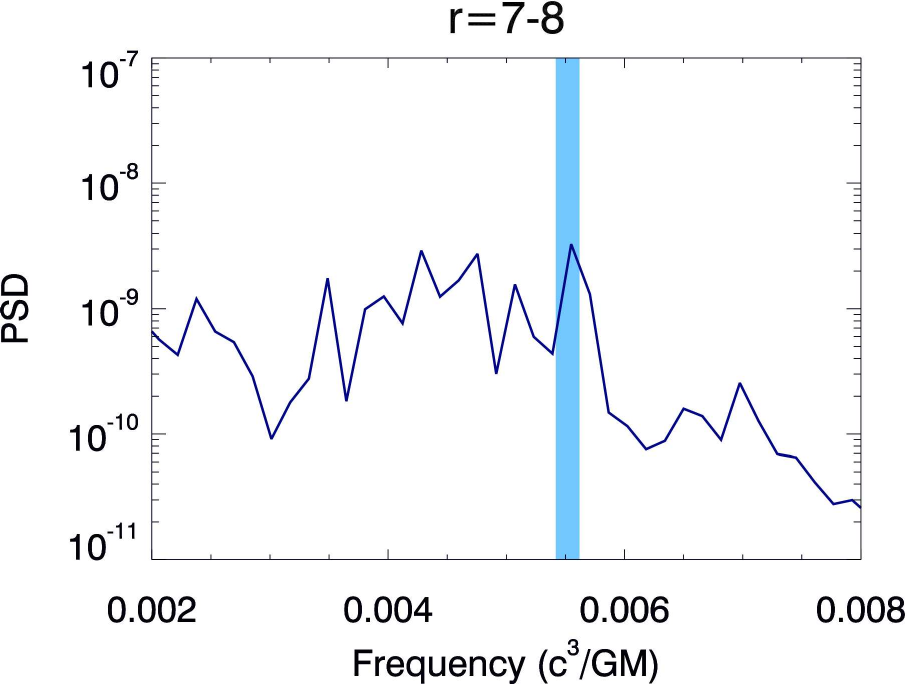}
\includegraphics[type=pdf,ext=.pdf,read=.pdf,width=0.329\textwidth]{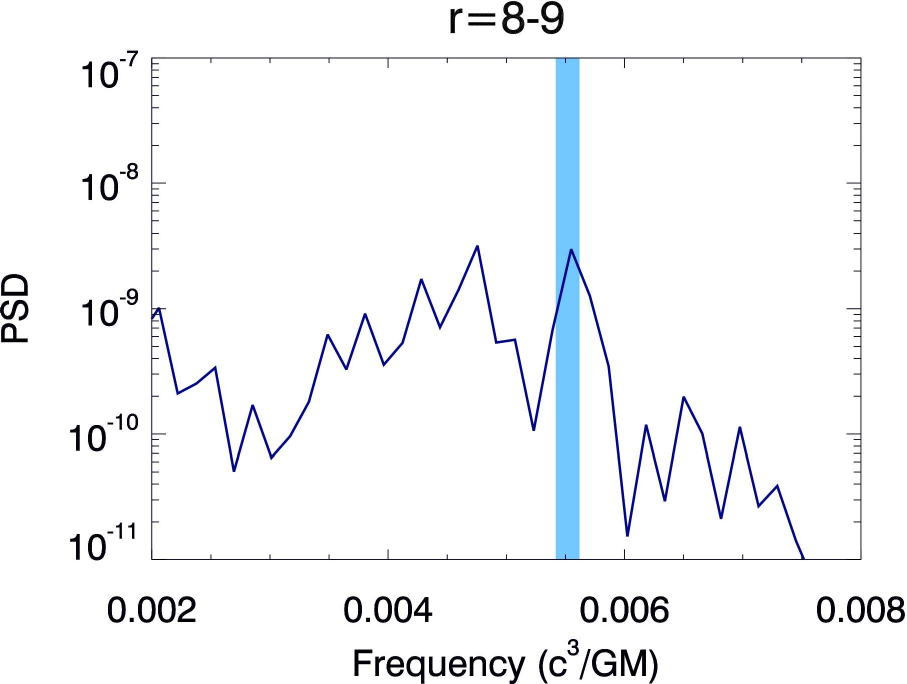}
\\[0.75em]
\includegraphics[type=pdf,ext=.pdf,read=.pdf,width=0.329\textwidth]{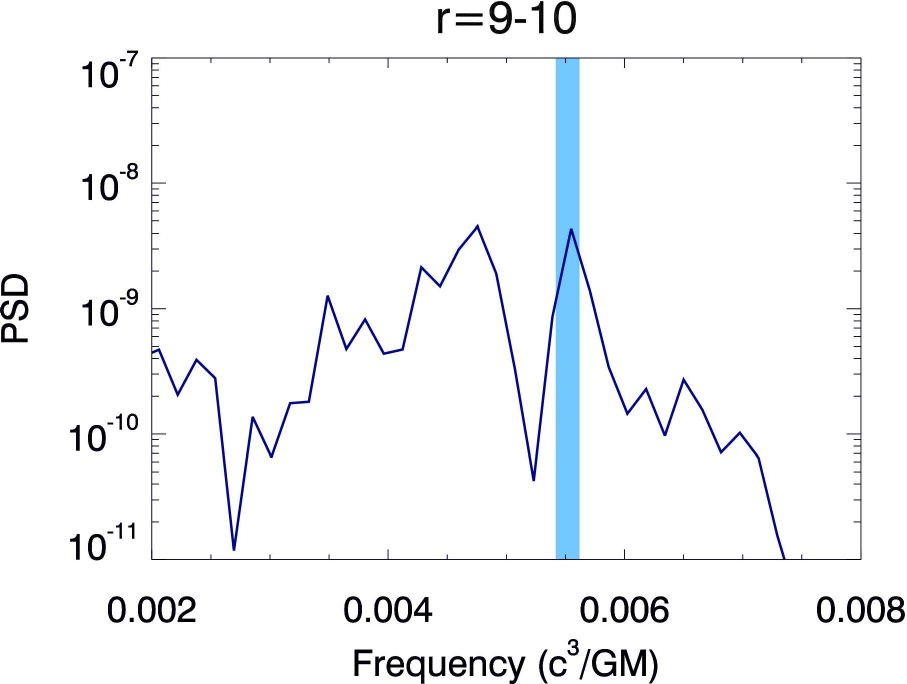}
\includegraphics[type=pdf,ext=.pdf,read=.pdf,width=0.329\textwidth]{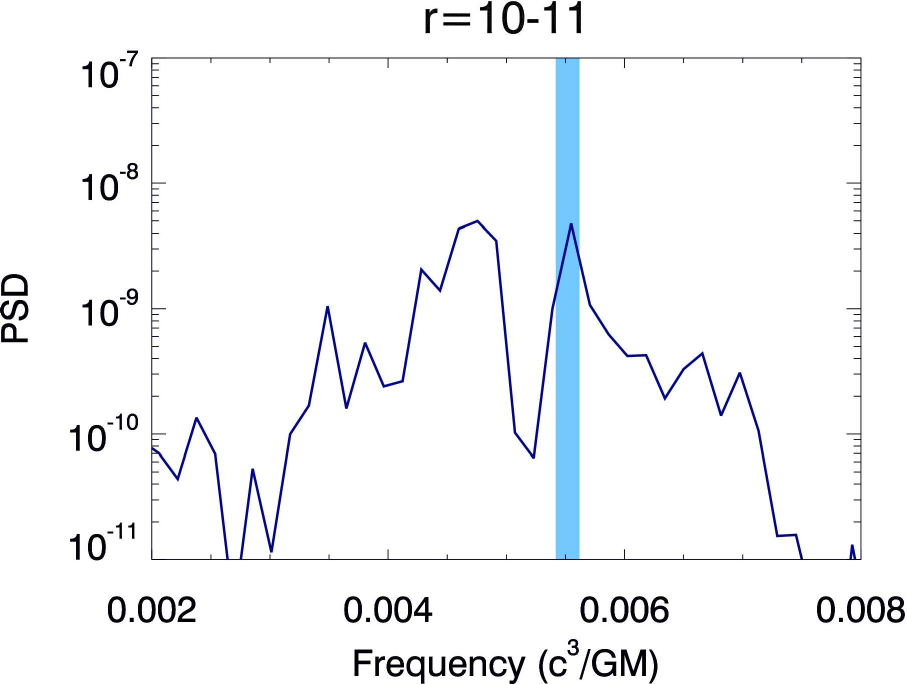}
\includegraphics[type=pdf,ext=.pdf,read=.pdf,width=0.329\textwidth]{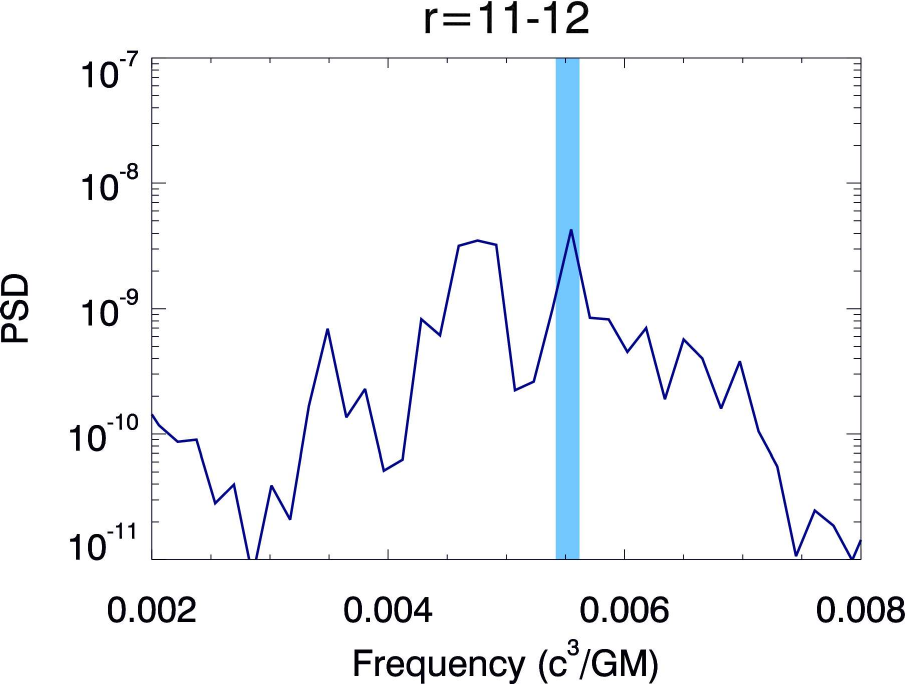}
\\[0.75em]
\includegraphics[type=pdf,ext=.pdf,read=.pdf,width=0.329\textwidth]{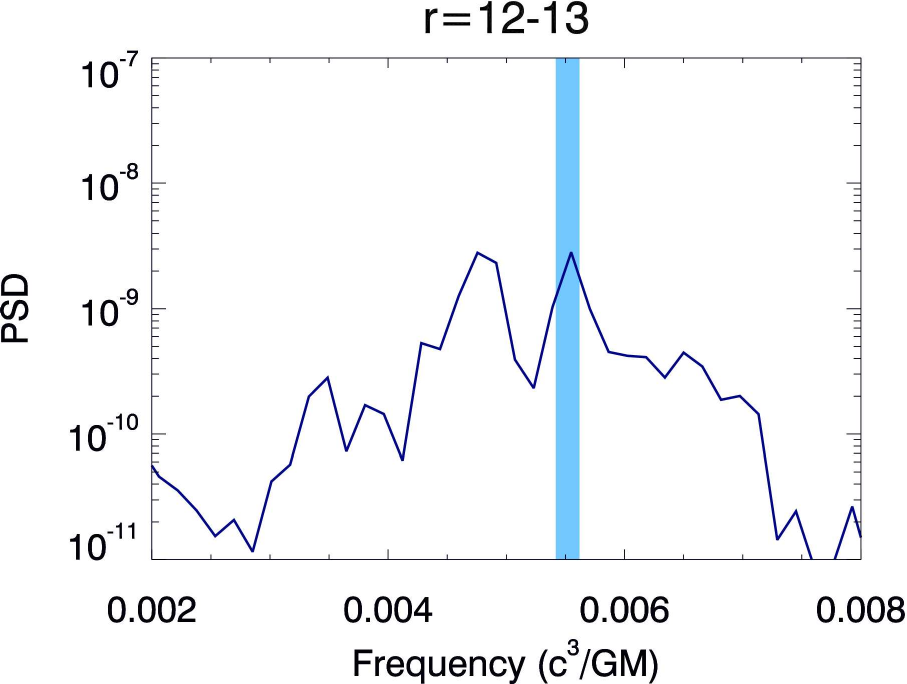}
\includegraphics[type=pdf,ext=.pdf,read=.pdf,width=0.329\textwidth]{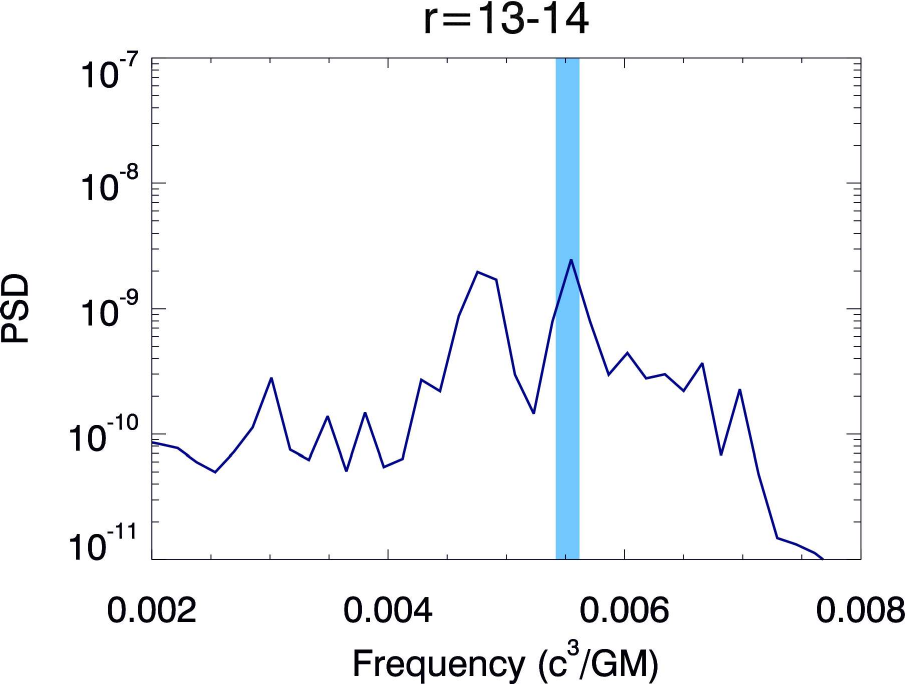}
\includegraphics[type=pdf,ext=.pdf,read=.pdf,width=0.329\textwidth]{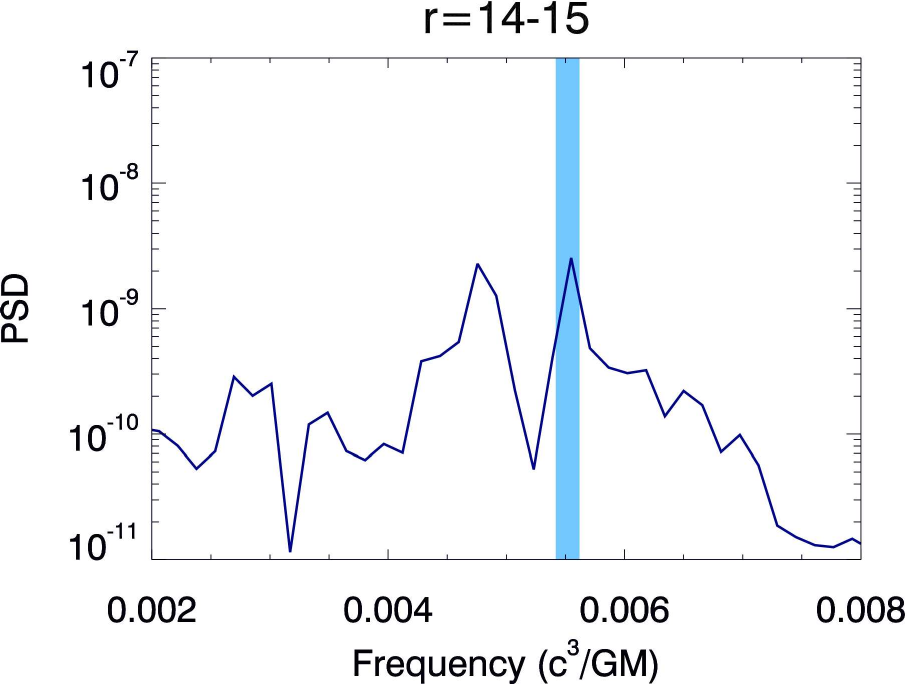}
\\[0.75em]
\includegraphics[type=pdf,ext=.pdf,read=.pdf,width=0.329\textwidth]{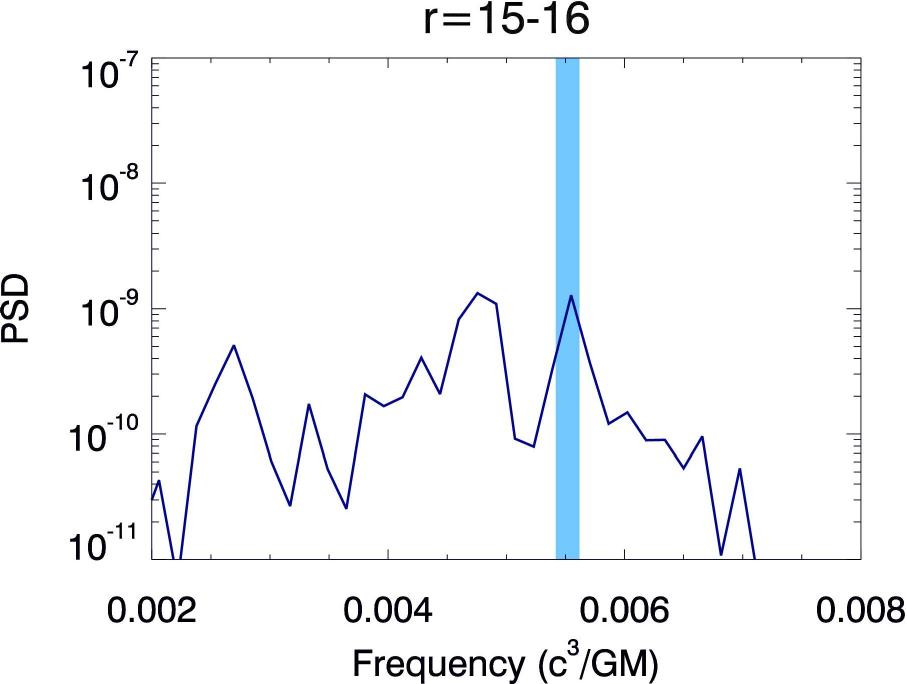}
\includegraphics[type=pdf,ext=.pdf,read=.pdf,width=0.329\textwidth]{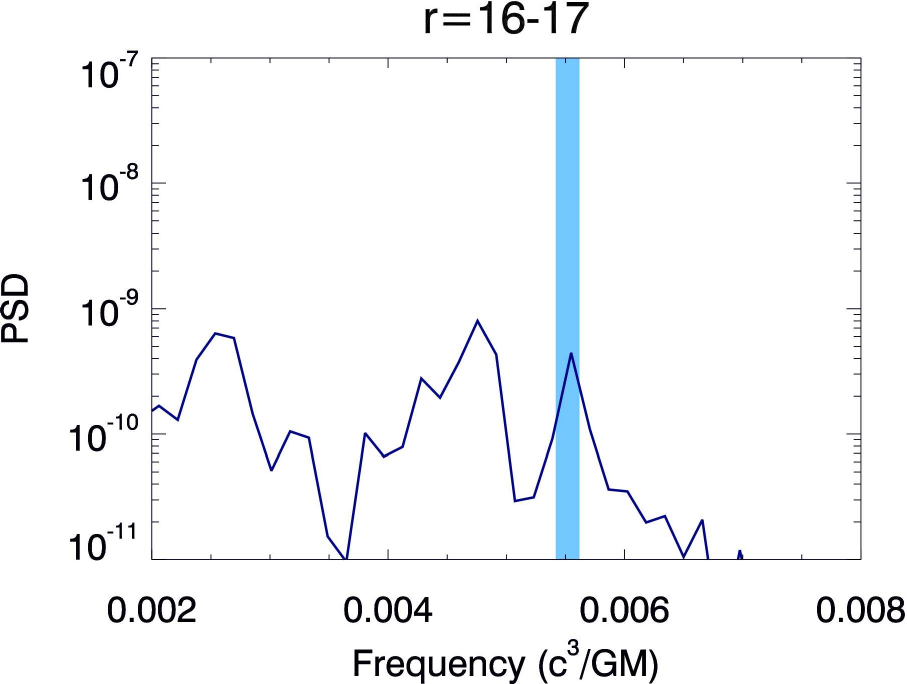}
\includegraphics[type=pdf,ext=.pdf,read=.pdf,width=0.329\textwidth]{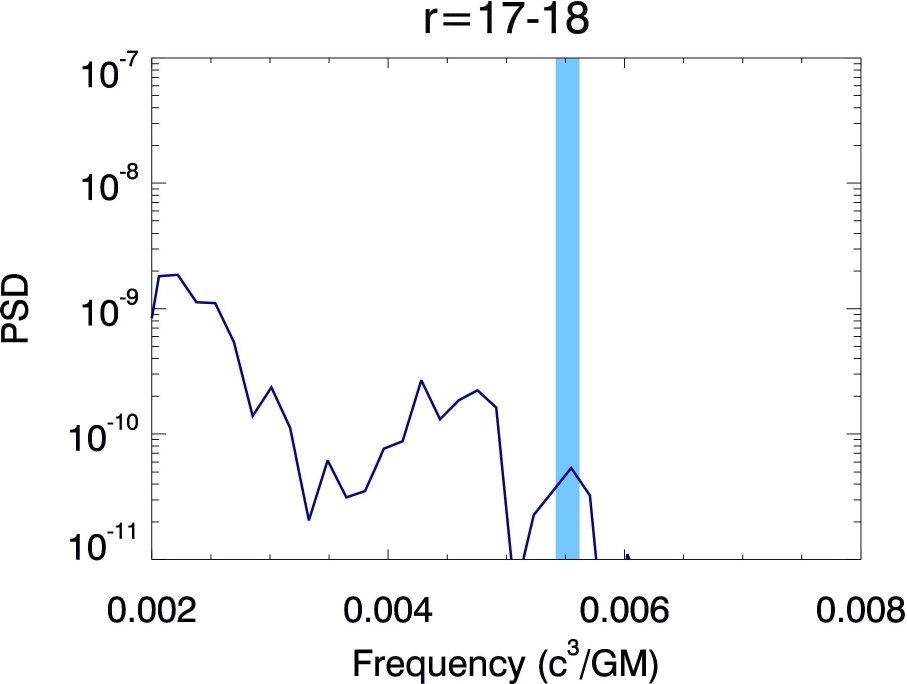}
\caption[Linear superposition of $v_{\rm r}$]{Linear superposition of the midplane PSDs of radial velocity in model A0.1 for a set of discrete radial bins.  The thick vertical line represents the position of the maximum radial epicyclic frequency, $\nu_{\rm max}$.  The peaks near $\nu_{\rm max}$, which are present throughout the radial range $r=9-16~GM/c^2$, indicate the trapped modes and the waves they produce.}
\label{fig:lin_vr}
\end{figure*}

To explore the radial dependence of spectral power in more detail, we show in Figure \ref{fig:lin_vr} a set of PSDs of the radial velocity in A0.1 linearly added in radius bins of width $\Delta r = 1~{\rm GM/c^2}$.
The vertical bar indicates the position of $\nu_{\rm max}$ in each plot.
At the smallest radii ($r=6-7~{\rm GM/c^2}$), there is sufficient noise in the bin to preclude simple identification of the aforementioned spikes.
While the background noise is slightly diminished at $r=7-9~{\rm GM/c^2}$, the spikes are only easily seen by $r \ge 9~{\rm GM/c^2}$.
Interestingly, the spikes maintain a power level that is constant to within a factor of two from $r=9-15~{\rm GM/c^2}$, and they are seen to peak at or just below $\nu_{\rm max}$ at these radii.
In the last three radius bins, we see that the amplitude of the signal finally diminishes as lower frequency noise begins to creep in at outer disk radii.

\begin{figure}[t]
\includegraphics[type=pdf,ext=.pdf,read=.pdf,width=0.48\textwidth]{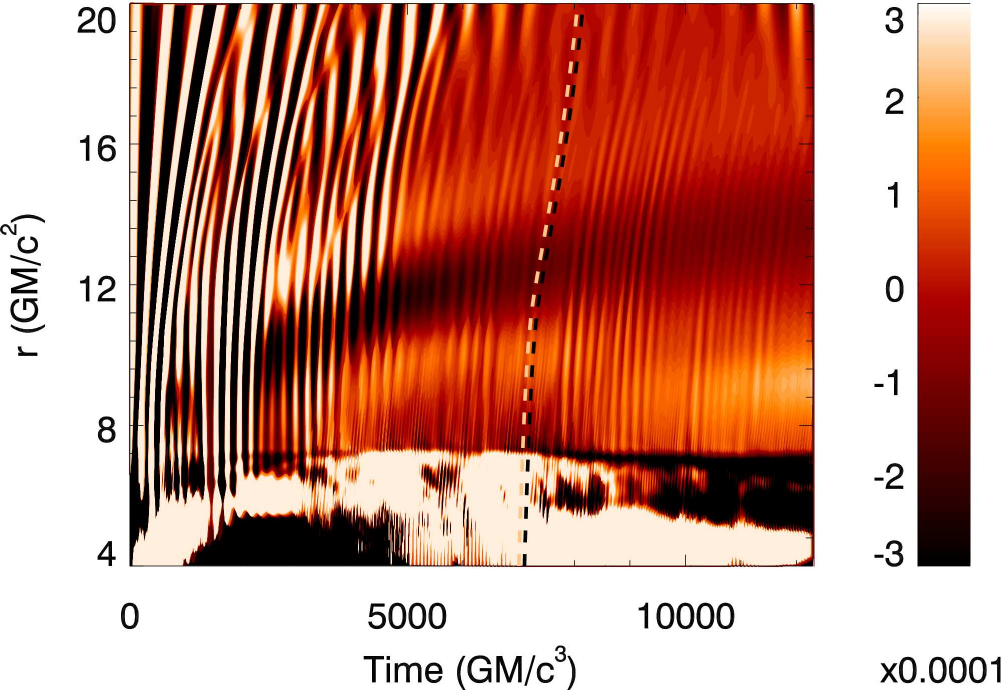}
\caption[Midplane radial velocity for A0.1]{The deviation in midplane radial velocity ($\Delta v_{\rm r} \equiv v_{\rm r} - {\bar v}_{\rm r}$) for A0.1 as a function of radius and time.  The linear color table extends from $v_{\rm r} = -0.0003c~{\rm (dark)~to}~v_{\rm r}=0.0003c~{\rm (light)}$, where positive radial velocities point radially outward in the disk.  The bi-colored dashed line, which can be arbitrarily shifted horizontally, represents the approximate path of a wave moving radially through the disk at the local sound speed.  That the velocity features run parallel to this line suggests that they are waves propagating radially outward from a region located near $r \sim r_{\rm max}$.}
\label{fig:a1em1vrmid}
\end{figure}

Although PSDs are invaluable for locating such features, we must appeal to alternate methods to determine the radius range from which these spikes emanate.
Figure \ref{fig:a1em1vrmid} shows for model A0.1 as a function of both time and radius the deviation in midplane radial velocity, defined as $\Delta v_{\rm r} \equiv v_{\rm r} - {\bar v}_{\rm r}$.
Shown as a bi-colored dashed line is a characteristic outward wave propagation path, derived from the local sound speeds averaged over the entire simulation time.
In this plot, we see that multiple streams form a combtooth pattern that, for $t > t_{\rm relax}$, originates at the plateau near $r \sim 7.5~{\rm GM/c^2}$, runs approximately parallel to the dashed line, and finally fades from view by $r \sim 18~{\rm GM/c^2}$.
That the pattern runs roughly parallel to the dashed line suggests that these are waves moving radially outward from their point of origin in the inner disk.
Where these waves vanish near the top of Figure \ref{fig:a1em1vrmid} is near where Figures \ref{fig:a1em1psds} and \ref{fig:lin_vr} suggest that a signal associated with the outer grid boundary begins to manifest itself \citepalias[such features were also noted in][]{2008arXiv0805.2950R}.
To avoid contamination from such boundary effects, we restrict our analysis to radii inward of $r \sim 18~{\rm GM/c^2}$, a range that still encompasses a large portion of the disk external to $r_{\rm ISCO}$ and $r_{\rm max}$.

We propose that the spikes and associated signals seen in Figures \ref{fig:a1em1psds}-\ref{fig:a1em1vrmid} are the natural result of viscosity-induced trapped oscillation modes, such as those described by \citet{2000ApJ...537..922O}.
In this framework, viscosity provides a mechanism by which the rotational velocity of the disk can be tapped by the orthogonal velocity components, leading to a driven, trapped mode (or modes).
The narrow frequency range of the spikes, located most prominently at or just below $\nu_{\rm max}$, is what would be expected from a trapped g-mode, although in practice distinguishing between inner p-modes and g-modes is not trivial.
For example, viscous disks characteristically share power locally between orthogonal velocity components, so one cannot simply assume that the presence of a signal in $v_{\rm z}$ indicates a g-mode.
Figures \ref{fig:a1em1psds} and \ref{fig:lin_vr} in principle could be used to identify the exact radial range for these modes, but broadband noise at the radii of interest ($r \sim 6-9~{\rm GM/c^2}$) makes it difficult to cleanly separate the two distinct mode trapping regions.
Moreover, it is clear from comparing Figures \ref{fig:a1em1psds} and \ref{fig:a1em1vrmid}, for example, that one cannot rely upon PSDs to distinguish proper trapped modes from induced wave motions.
In fact, our only strong constraint on the location of these modes comes from Figure  \ref{fig:a1em1vrmid}, which shows that the modes themselves are not present exterior to $r_{\rm max}$.
Given that trapped p- and g-modes are apparently indistinguishable in our simulations, we will thus refer to these features generically as ``trapped modes'' for the remainder of this discussion.

A combined analysis of Figures \ref{fig:a1em1psds} - \ref{fig:a1em1vrmid} clearly illustrates that the influence of these trapped modes in A0.1 is not restricted to that portion of the disk within the trapped region.
Figures \ref{fig:a1em1psds} and \ref{fig:lin_vr}, for example, demonstrate that the disk contains significant power near $\nu_{\rm max}$ for $r \ge r_{\rm max}$.
Figure \ref{fig:a1em1vrmid} shows that this power exterior to $r_{\rm max}$ is in the form of outward propagating waves.
That these waves are related to the trapped modes is evident from their discrete frequencies, which are identical to those of the trapped modes.
Still, it is challenging to identify exactly how the modes transfer their energy to these waves since both trapped g-modes and inner p-modes are formally evanescent at frequencies greater than the radial epicyclic frequency.
Moreover, there is no obvious indication that the signal loses power across the radial epicyclic boundary, as might be expected for mode leakage.
One simple plausible explanation is that the modes excite radial waves, which are non-vanishing for all frequencies greater than the radial epicyclic frequency \citep[see][for example]{1993ApJ...409..360L}.
One could imagine predominantly vertical trapped g-modes, for example, exciting through viscous action radial waves that then propagate freely outside of the radial epicyclic boundary.
Another possible explanation is that the trapped modes tunnel through the finite evanescent region, which is bounded by the radial epicyclic and orbital frequencies in the case of the axisymmetric fundamental p-mode. 
Assuming that such waves do not decay appreciably in the evanescent zone, they could emerge as radial p-modes in regions of the outer disk for which the local orbital frequency is less than the original trapped mode frequency.
Whatever the mechanism, the ultimate result is that these waves retain the frequency signature of the trapped modes as they move through the disk.
Eventually, these waves become lost in the artificial noise generated by the outer grid boundary, but not before intersecting a significant fraction of the disk body.

\begin{figure*}
\includegraphics[type=pdf,ext=.pdf,read=.pdf,width=0.54\textwidth]{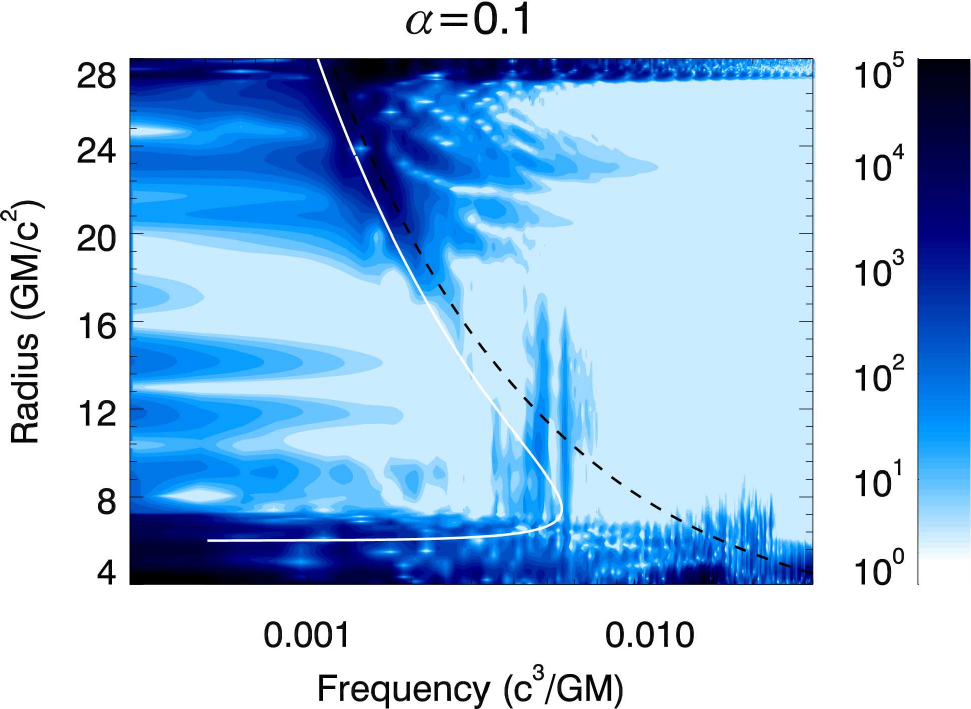}
\includegraphics[type=pdf,ext=.pdf,read=.pdf,width=0.54\textwidth]{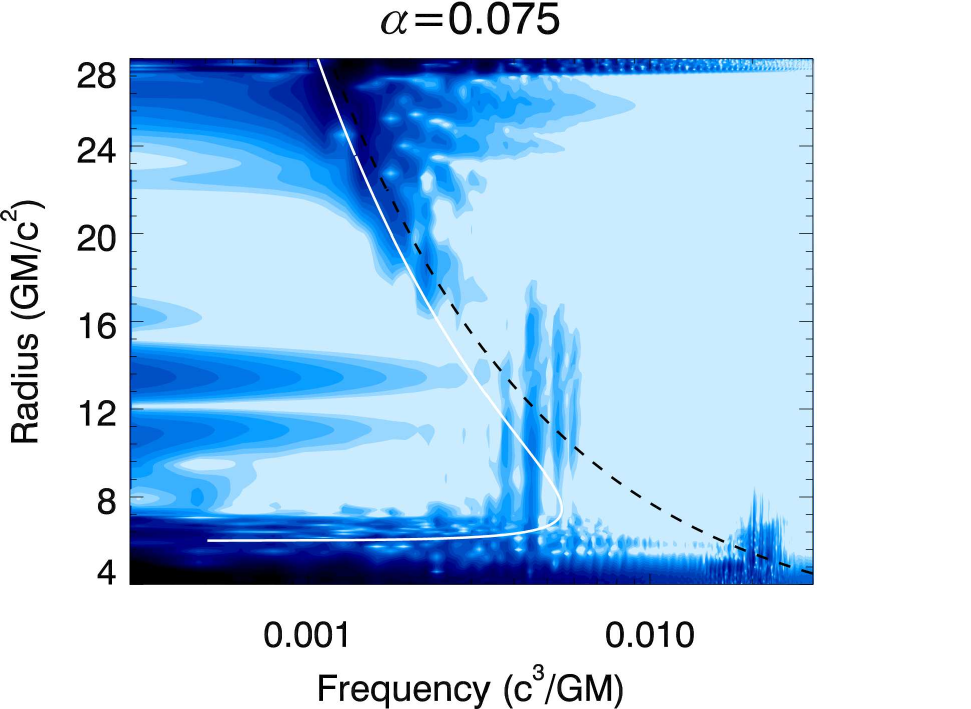}
\\[0.75em]
\includegraphics[type=pdf,ext=.pdf,read=.pdf,width=0.54\textwidth]{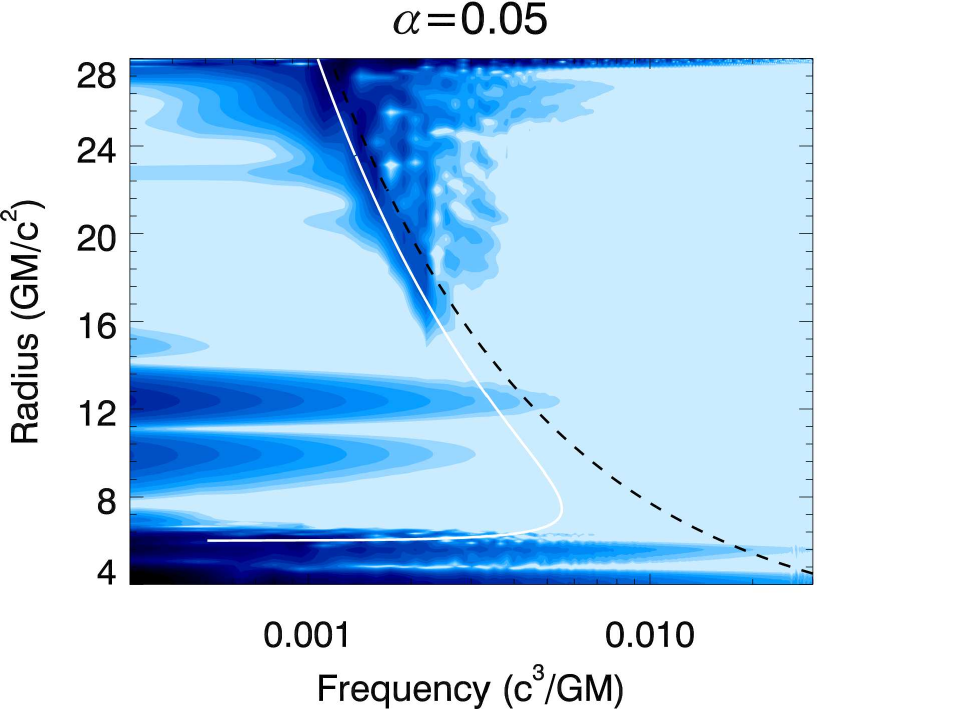}
\includegraphics[type=pdf,ext=.pdf,read=.pdf,width=0.54\textwidth]{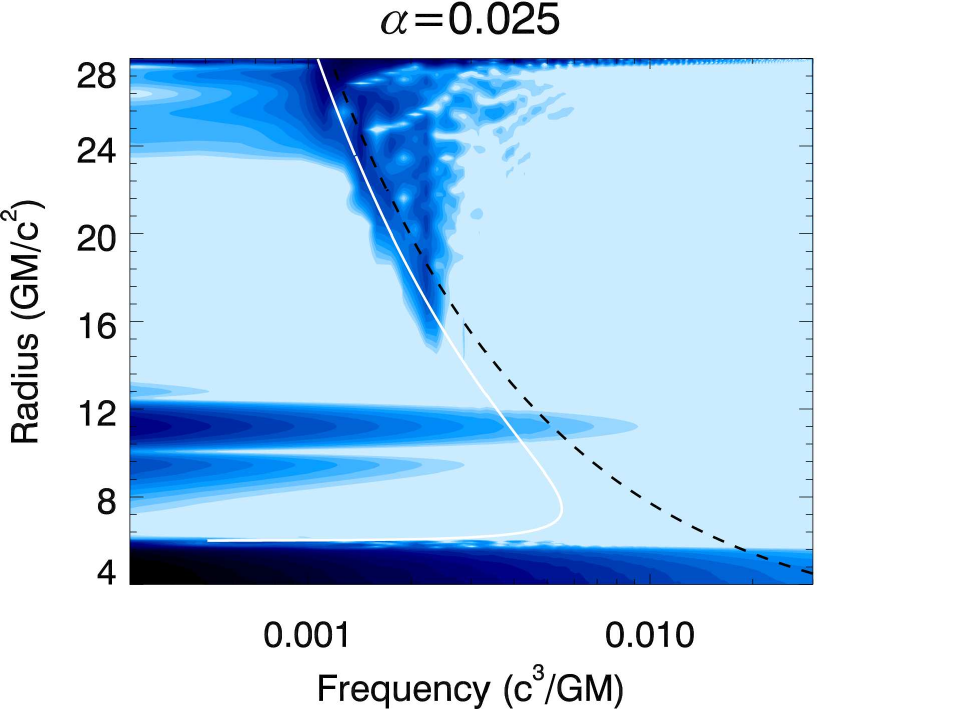}
\\[0.75em]
\includegraphics[type=pdf,ext=.pdf,read=.pdf,width=0.54\textwidth]{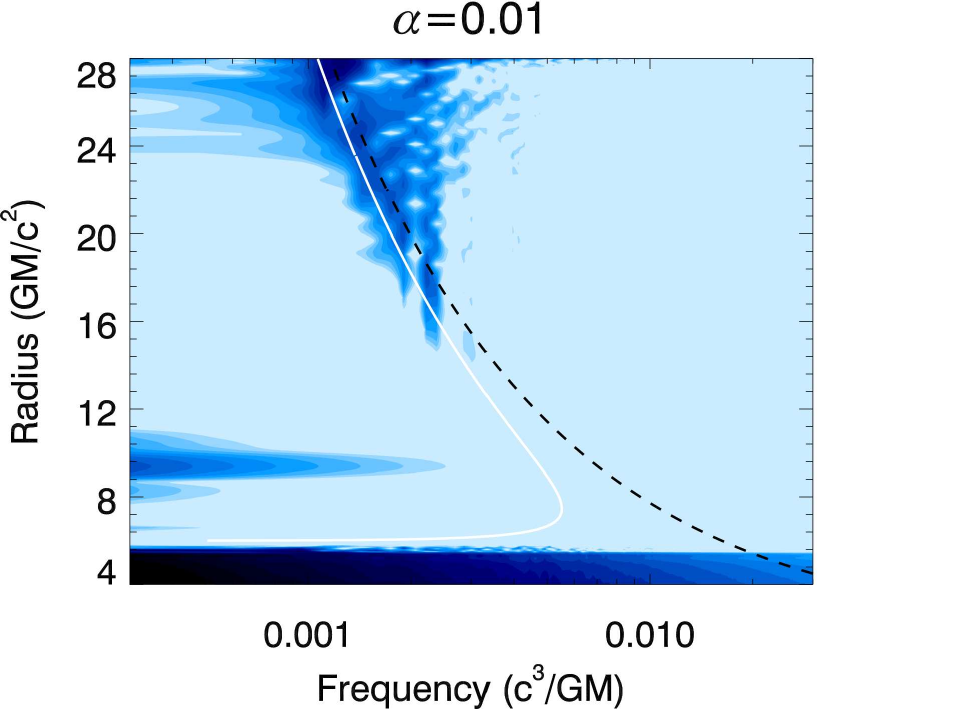}
\includegraphics[type=pdf,ext=.pdf,read=.pdf,width=0.54\textwidth]{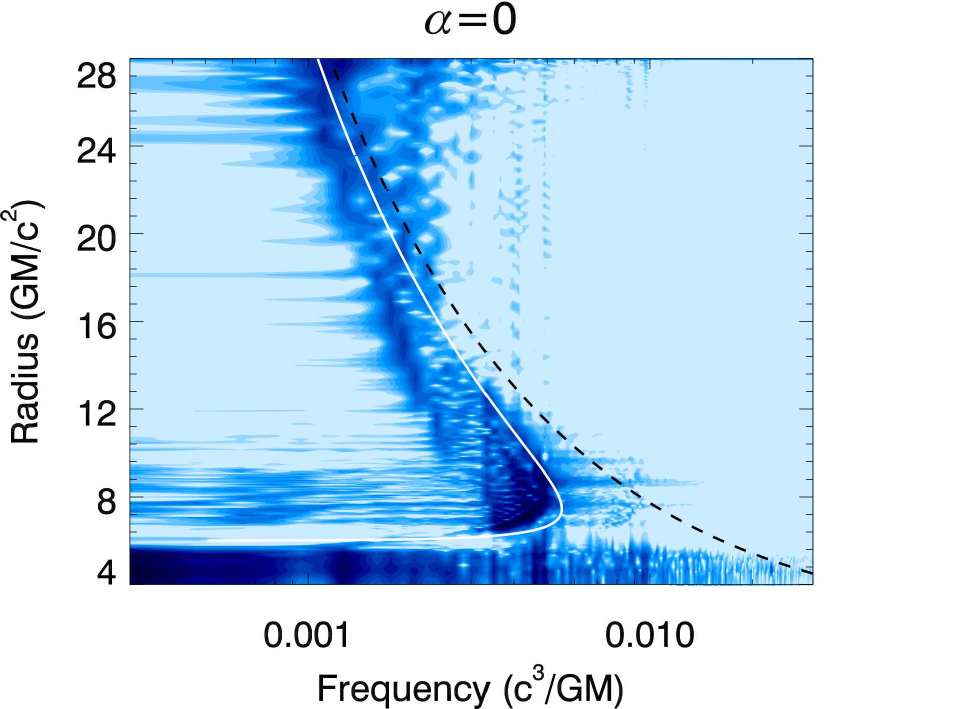}
\caption[Midplane PSDs of radial velocity]{Midplane PSDs of radial velocity for a range of model viscosities.  Also shown are the radial epicyclic frequency (solid) and orbital frequency (dashed) for comparison.  The logarithmic colorbars span five orders of magnitude and are normalized in magnitude across all plots in this figure to facilitate cross-comparison.  Here, models A0.1 and A0.075 clearly show trapped modes and waves, and A0 features a trapped g-mode.}
\label{fig:vrpsds}
\end{figure*}

Before moving on to our discussion of other disk models, it is worth discussing the slight indications of a signal visible near $\nu_{\rm max}$ for $r < r_{\rm ISCO}$ in the velocity PSDs in Figure \ref{fig:a1em1psds}.
This represents leakage of the trapped modes down through the ISCO into a narrow accretion stream that leaves the grid, similar to that noted in \citetalias{2008arXiv0805.2950R}.
There is no such leaked signal visible in the density or pressure PSDs because these quantities are significantly reduced in magnitude radially inward from $r_{\rm max}$ through $r_{\rm ISCO}$.
The midplane density, for example, is over two orders of magnitude smaller in the accretion stream than at $r \ge r_{\rm max}$. 
Any signal proportional to the local density would thus have four orders of magnitude less power in the PSD at the innermost radii than at $r \ge r_{\rm max}$.
This typically pushes such a signal below the range of our colorbar, and we have confirmed that such signals are indeed present, but weak.

\begin{figure*}
\includegraphics[type=pdf,ext=.pdf,read=.pdf,width=0.54\textwidth]{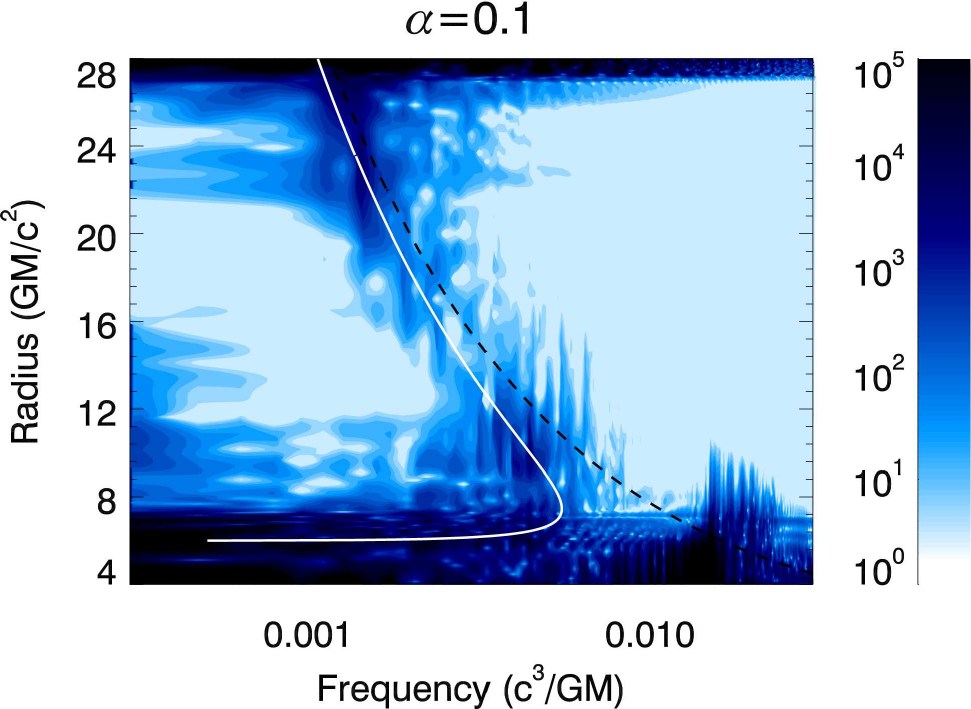}
\includegraphics[type=pdf,ext=.pdf,read=.pdf,width=0.54\textwidth]{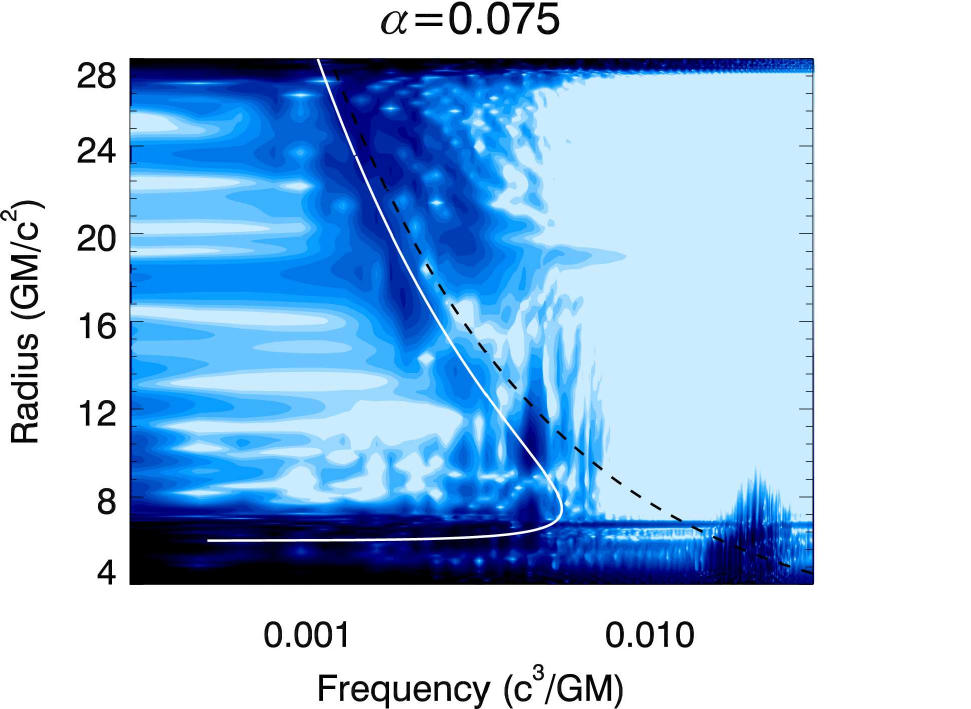}
\\[0.75em]
\includegraphics[type=pdf,ext=.pdf,read=.pdf,width=0.54\textwidth]{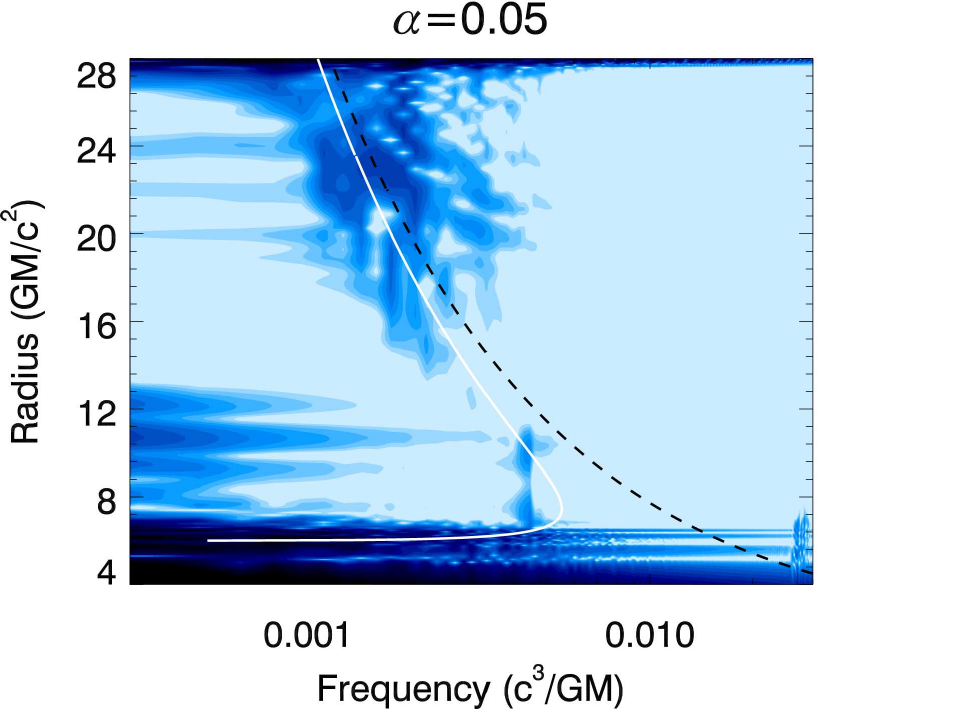}
\includegraphics[type=pdf,ext=.pdf,read=.pdf,width=0.54\textwidth]{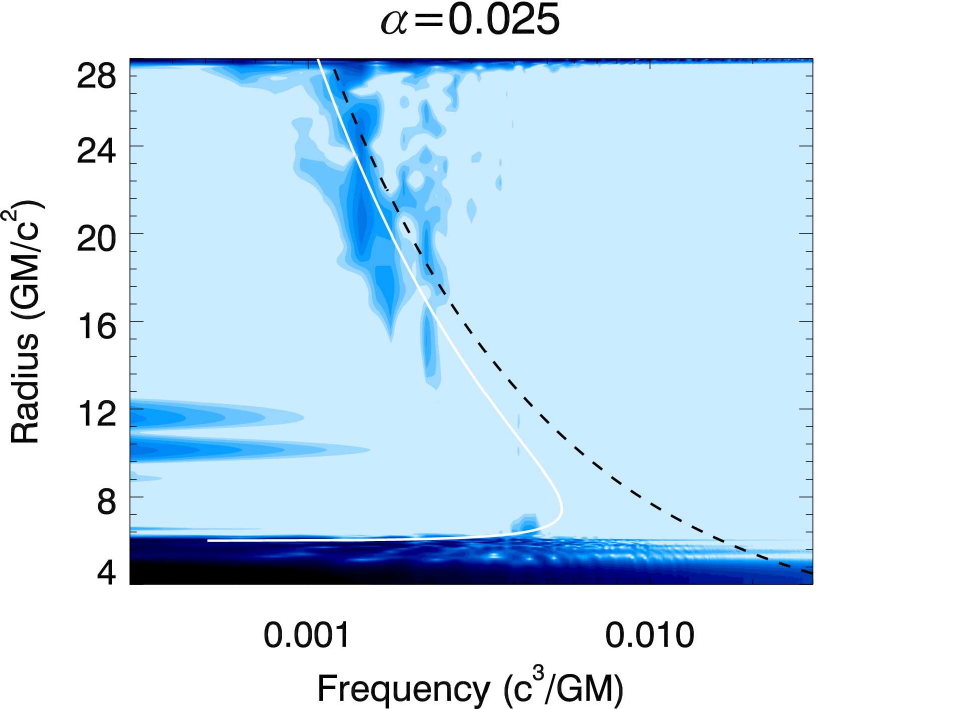}
\\[0.75em]
\includegraphics[type=pdf,ext=.pdf,read=.pdf,width=0.54\textwidth]{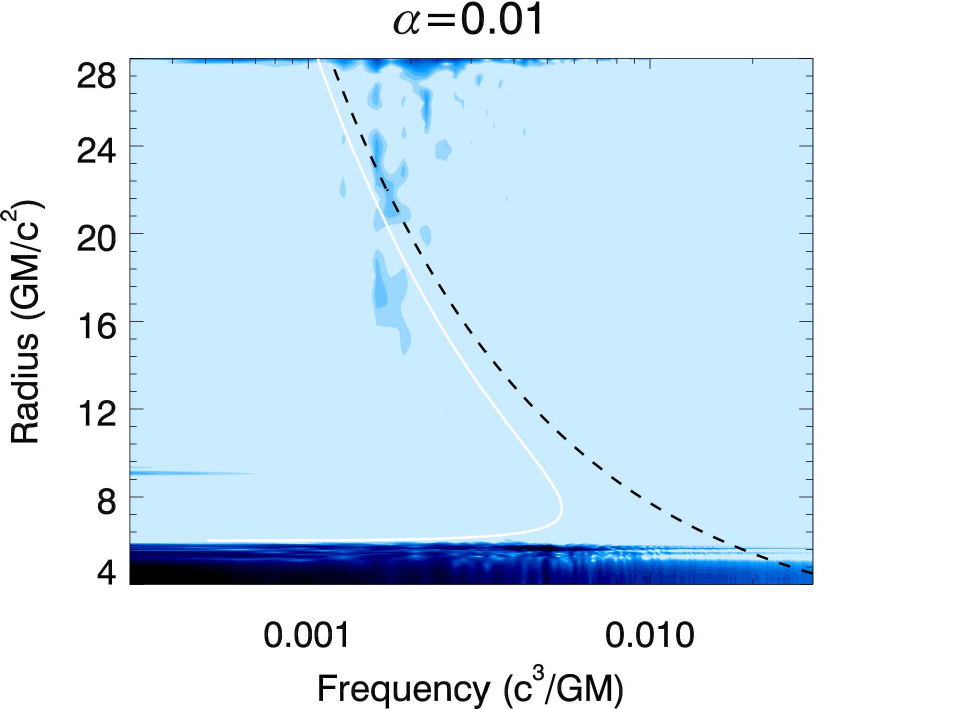}
\includegraphics[type=pdf,ext=.pdf,read=.pdf,width=0.54\textwidth]{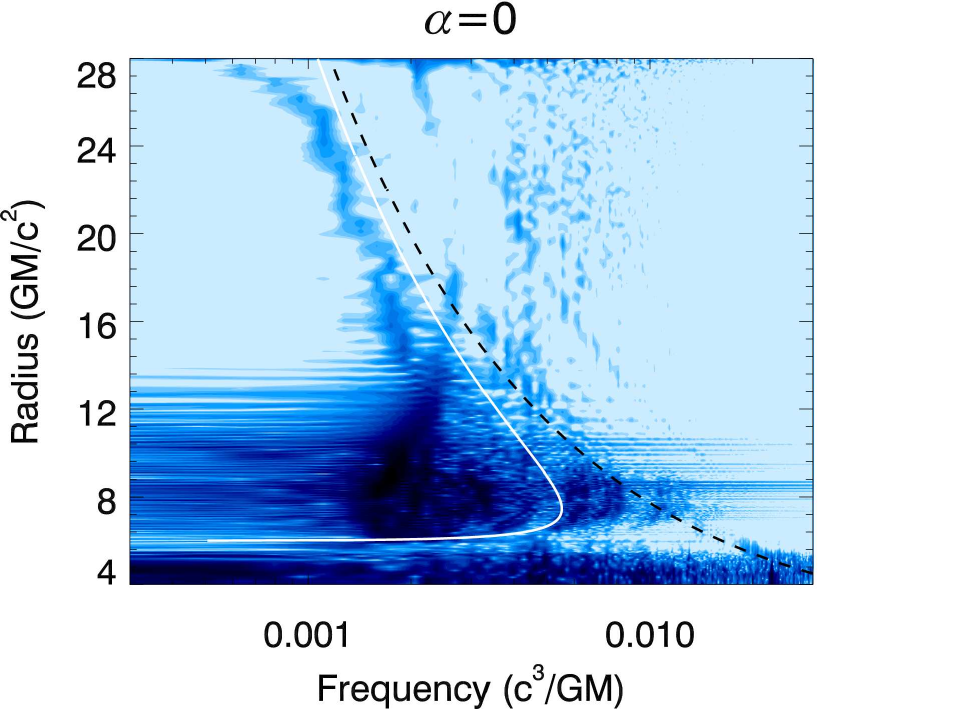}
\caption[Midplane PSDs of vertical velocity]{Midplane PSDs of vertical velocity for a range of model viscosities.  Also shown are the radial epicyclic frequency (solid) and orbital frequency (dashed) for comparison.  The logarithmic colorbars span five orders of magnitude and are normalized in magnitude across all plots in this figure to facilitate cross-comparison.  Here, models A0.1 and A0.075 clearly show trapped modes and waves, A0.05 shows evidence for a trapped mode, and A0 features a trapped g-mode.}
\label{fig:vzpsds}
\end{figure*}

The presence and influence of trapped modes are not exclusive to model A0.1, but are seen for other viscosities as well.
Figure \ref{fig:vrpsds} shows the radial velocity PSDs and Figure \ref{fig:vzpsds} the vertical velocity PSDs for A0.1, A0, and the four intermediate viscous disk models.
The color bars in Figures \ref{fig:vrpsds} and \ref{fig:vzpsds} are each normalized so that we can correctly cross-compare magnitudes across different models in each figure.
First, the basic trapped mode features are present in both velocity components for viscous models with $\alpha \ge 0.075$.
While the exact positions and number of detectable spikes differ in detail between A0.1 and A0.075, the proximity of these spikes to $\nu_{\rm max}$ still reflects the basic orbital parameters of our model disks.
Moving to lower viscosities, model A0.05 shows an identifiable signal in vertical velocity that appears to be spatially bounded by the radial epicyclic frequency.
There is no analogous signal in radial velocity, however, even if we examine a range in power below that shown in Figure \ref{fig:vrpsds}.
Similarly, model A0.025 shows the hint of a signal in vertical velocity for $r < r_{\rm max}$, but absolutely no signal in the radial velocity.
Finally, Model A0.01 features no significant signal for either velocity component or any choice of PSD range.

One might ask whether these trapped modes are identical to the trapped g-modes seen in A0 or whether these are truly the distinct viscosity-induced trapped modes described by \citet{2000ApJ...537..922O}.
Looking again at Figure \ref{fig:kdecayrates}, it is clear that the initial disk perturbation energy is damped out more rapidly at early times in the viscous models than in A0, thus depriving alpha-disks of much of the initial energy available to inviscid disks.
Moreover, A0 shows a long-term downward trend upon which is superposed a high-frequency signal indicative of the trapped g-mode.
Model A0.1, on the other hand, has no such obvious long-term trend, suggesting that viscous and boundary losses are offset by ongoing energy input.
In the alpha-disk models, this input energy in fact stems from the viscous coupling of disk rotational velocity to radial and vertical motions; a channel unavailable to inviscid disks.
Although the peaks and valleys in Figure \ref{fig:kdecayrates} illustrate that, for model A0.1, this process has not achieved a steady-state on timescales much shorter than the total simulation time, the overall energy profile demonstrates that the viscous method of generating persistent trapped modes is distinct from that of an inviscid disk.
Figure \ref{fig:kdecayrates} also shows models A0.05 and A0.01, both of which evolve to a lower value of $K$.
This is not surprising since their trapped mode signals are weaker or, in the case of A0.01, undetectable, suggesting that the energy resupply is not as efficient as for higher viscosities.
Interestingly, model A0.01 does show an increase in $K$ near the very end of its evolution, but we cannot definitively claim this as evidence for the development of trapped modes without extending the simulation in time.

\begin{figure*}
\includegraphics[type=pdf,ext=.pdf,read=.pdf,width=0.54\textwidth]{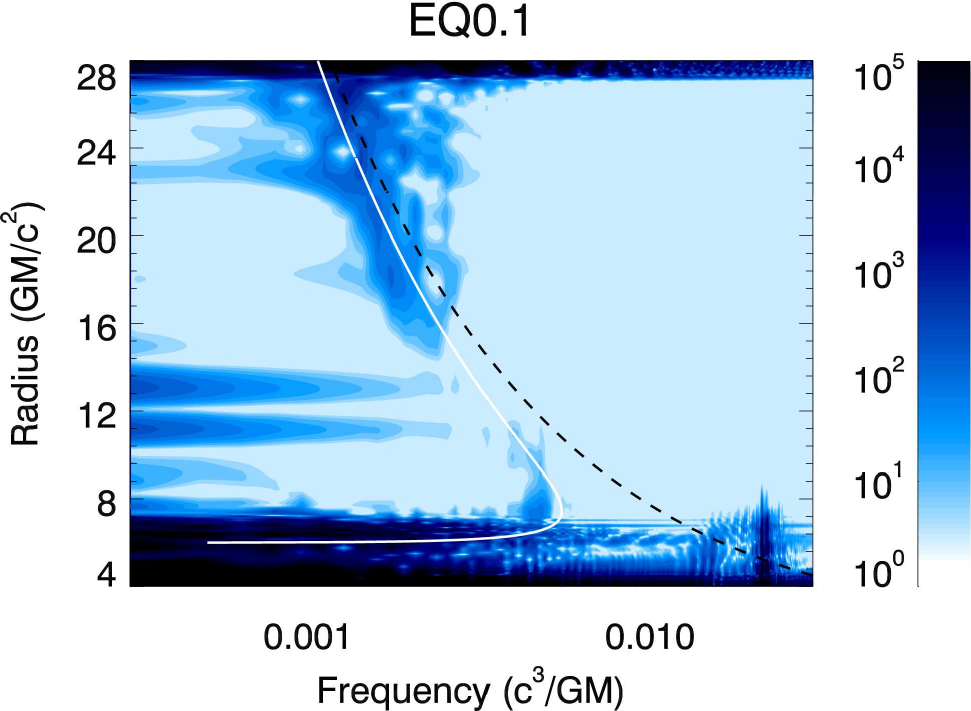}
\includegraphics[type=pdf,ext=.pdf,read=.pdf,width=0.54\textwidth]{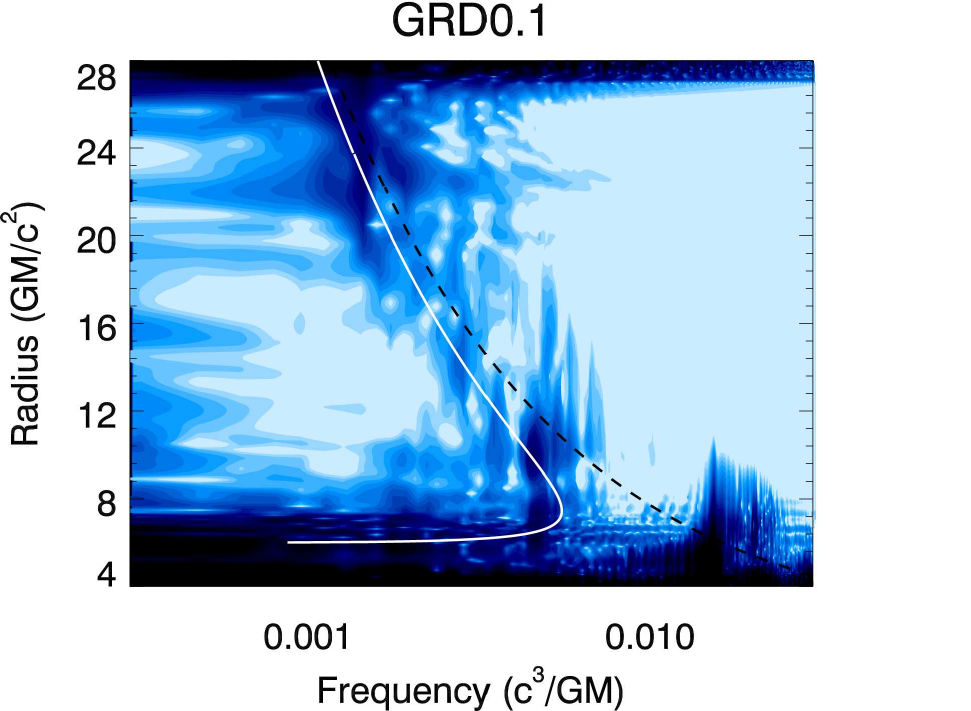}
\caption[Midplane PSDs of vertical velocity for test runs]{Midplane PSDs of vertical velocity for the two test simulations EQ0.1 and GRD0.1.  Also shown are the radial epicyclic frequency (solid) and orbital frequency (dashed) for comparison.  The logarithmic colorbars span five orders of magnitude.  EQ0.1 resembles one of the lower viscosity runs because it has a lower average temperature than A0.1.  GRD0.1, on the other hand, strongly resembles A0.1.}
\label{fig:testpsds}
\end{figure*}

Finally, we reiterate that the signal observed in our viscous models cannot be generated by unphysical computational phenomena.
Test models EQ0.1 and GRD0.1 were designed specifically to confirm that conditions such as the disk perturbation method and the computational grid boundary locations were not important factors in our simulations.
Figure \ref{fig:testpsds} shows the vertical velocity PSDs for the two test simulations.
EQ0.1 resembles one of the intermediate viscosity models with a weaker trapped mode signal than that of A0.1.
This is partly because the kinematic viscosity is lower in EQ0.1 than in A0.1.
Recall from Equation \ref{eq:alphavisc} that $\nu = \alpha c_{\rm s} H \propto c_{\rm s}^2$, which is in turn proportional to the disk temperature.
Since the equilibrium disk in model EQ0.1 does not collapse, it is not adiabatically heated and is subsequently cooler on average than the disk in model A0.1.
Empirically, we measure the average kinematic viscosity in the inner disk of model EQ0.1 to be $\sim 70\%$ that of A0.1.
We thus expect the trapped mode behavior of model EQ0.1 to fall roughly between that of A0.05 and A0.075, which is consistent with Figure \ref{fig:testpsds}.
Additionally, we note that the trapped mode signal in EQ0.1 should take longer than any other model to reach a given trapped mode amplitude since the seed disk perturbations are initially so small.
Model GRD0.1, on the other hand, reflects most of the significant characteristics of A0.1, showing that the trapped mode signal is not strongly dependent in the location of the grid boundaries.
Interestingly, this and all models feature a small region of high-frequency noise near $\nu \sim 0.01-0.02~{\rm c^3/GM}$.
We assume that this is associated with motions of material in the relatively diffuse accretion stream since the signal is comparable in frequency to the innermost orbital frequencies and does not appear in the pressure and density PSDs in Figure \ref{fig:a1em1psds}.

\section{Discussion}\label{sec:discussion}

\subsection{Comparison with Previous Results}

As mentioned in Section \ref{sec:introduction}, the work of \citet{1992PASJ...44..529H}, \citet{1995ApJ...441..354C,1996MNRAS.283..919M,1997MNRAS.286..358M}, and \citet{2008arXiv0805.0598M} has identified simulated accretion disk oscillations previously, and some of their work merits comparison here.
Their basic finding relevant to our work is that waves propagate through their simulated disks at frequencies $\sim \nu_{\rm max}$.
Specifically, \citet{1996MNRAS.283..919M} find waves in 1D vertically integrated disks for accretion rates $0.01 \dot{M}_{\rm Edd} \lesssim M \lesssim 0.25 \dot{M}_{\rm Edd}$ for a characteristic viscous parameter $0.2 \lesssim \alpha \lesssim 1$.
For our simulated accretion rates of $\dot{M}/\dot{M}_{\rm Edd} \sim 0.01$, we find that trapped modes and waves are easily identifiable for $\alpha = 0.075-0.1$, trapped modes only are marginally detectable at $\alpha = 0.05$, and all modes and waves are undetectable for lower viscosities.
Taking a detailed look at \citet{1996MNRAS.283..919M}, we see that they also detect a signal for $\alpha \ge 0.05$ in the case that $\dot{M}/\dot{M}_{\rm Edd} = 0.01$.
They claim no detection at $\alpha = 0.025$, where we too failed to detect even a convincing mode, and they simulate no lower viscosities.
The 2D simulations of \citet{1997MNRAS.286..358M} also find that the oscillations are favored for low accretion rates and high viscosities, but it is difficult to compare their work to our results since they simulate optically thick disks with accretion rates and viscosities characteristically higher than ours.
We note that these studies combine to suggest that oscillations are present for a broad variety of physical models, including both constant and alpha viscosities \citep{1996MNRAS.283..919M,1997MNRAS.286..358M}, and for both 2D and vertically integrated disks.

One interesting feature present in the aforementioned viscous disk simulations is a characteristic strong signal located in frequency very near $\nu_{\rm max}$.
Specifically, the higher-frequency spike seen in our models A0.1 and A0.075 (see Figures \ref{fig:a1em1psds}, \ref{fig:lin_vr}, and \ref{fig:vrpsds}) peaks at $\nu_{\rm max}$ to the accuracy of the PSD frequency resolution.
This is worth noting because analytic treatments of non-evanescent trapped g-modes and inner p-modes constrain them to have frequencies strictly less than $\nu_{\rm max}$ \citep[\eg][]{2000ApJ...537..922O,1992ApJ...393..697N}.
In fact, the lower-frequency spike seen in models A0.1 and A0.075 has exactly the expected characteristics of these predicted trapped modes, peaking just below $\nu_{\rm max}$.
Although the absence of a narrow signal in model A0 suggests that viscosity is partly responsible, no clear physical explanation of this high-frequency feature has yet been put forth, as \citet{2001PASJ...53....1K} also notes.

\subsection{Comparing $\alpha$-models to Full MHD}

Part of our motivation for conducting these simulations was to explore the differences between full MHD and viscous alpha-models.
To do this, we revisit one of the MHD simulations described in \citetalias{2008arXiv0805.2950R} and labeled ``MHD\_{\rm 1}''.
MHD\_{\rm 1} was a full 3D MHD simulation that utilized a computational framework and initial conditions similar to our alpha-models, extended axisymmetrically in the azimuthal dimension.
Additionally, MHD\_{\rm 1} featured initially weak poloidal magnetic field loops that threaded the disk.
As described in \citetalias{2008arXiv0805.2950R}, these fields are amplified by the MRI and drive turbulence which in turn provides a natural means for accretion to take place. 
The physical domain of MHD\_{\rm 1} covered $r \in (4r_{\rm g},16r_{\rm g})$, $z \in (-3r_{\rm g},3r_{\rm g})$,, and $\phi \in (0,\pi/6)$, and it was run for over three times the total simulation time of our alpha-disks.

\begin{figure*}
\includegraphics[type=pdf,ext=.pdf,read=.pdf,width=0.5\textwidth]{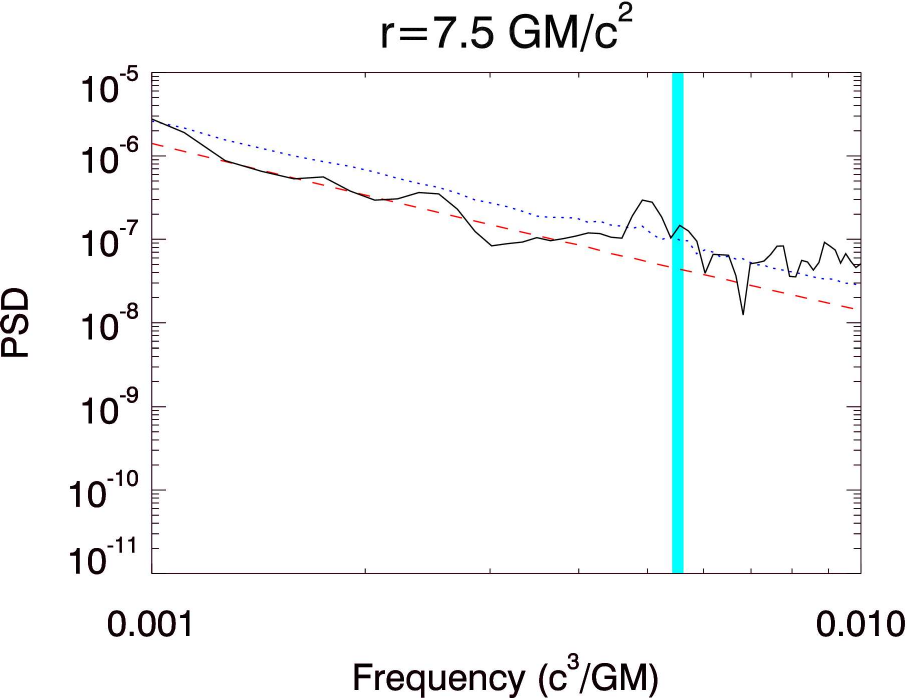}
\includegraphics[type=pdf,ext=.pdf,read=.pdf,width=0.5\textwidth]{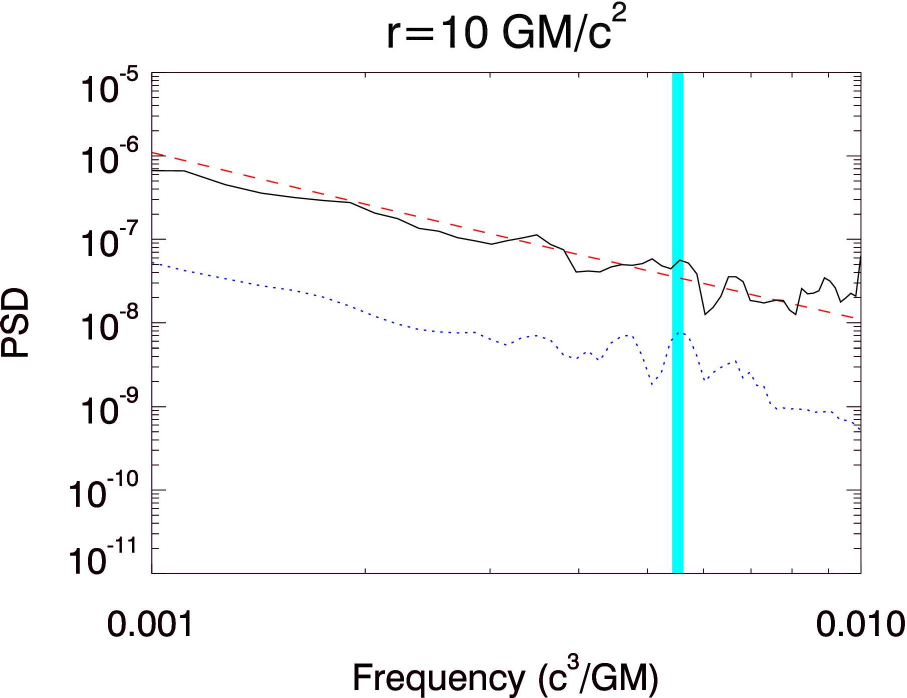}
\\[0.75em]
\includegraphics[type=pdf,ext=.pdf,read=.pdf,width=0.5\textwidth]{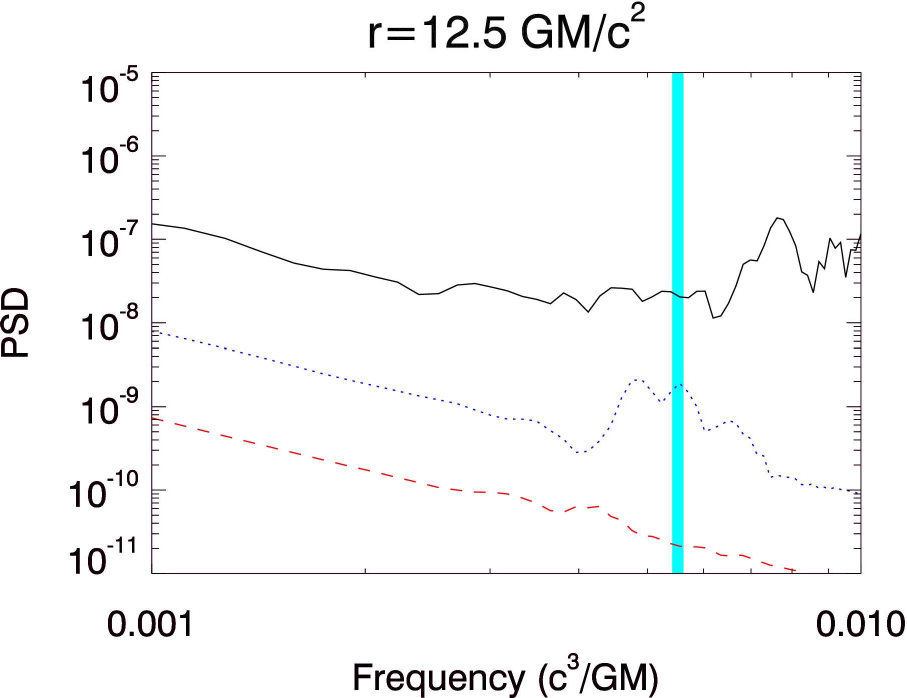}
\includegraphics[type=pdf,ext=.pdf,read=.pdf,width=0.5\textwidth]{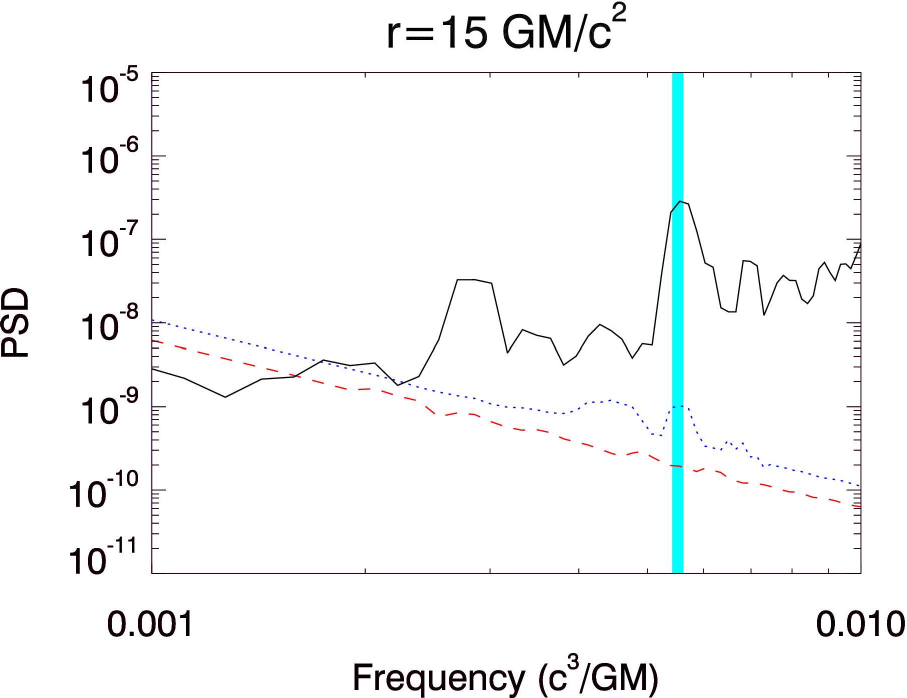}
\caption[Comparison of pressure]{Midplane PSDs of (decay-corrected) pressure, summed over radial ranges $\Delta r = 0.5r_{\rm g}$ and each centered at the listed radius.  Shown are models MHD\_{\rm 1} (solid), A0.1 (dotted), and A0.01 (dashed).  The vertical line indicates the position of the maximum radial epicyclic frequency.  In all cases, the signals of trapped modes in our model viscous disks would fall at least an order of magnitude below the noise level in MHD\_{\rm 1}.}
\label{fig:pcomp}
\end{figure*}

\begin{figure*}
\includegraphics[type=pdf,ext=.pdf,read=.pdf,width=0.5\textwidth]{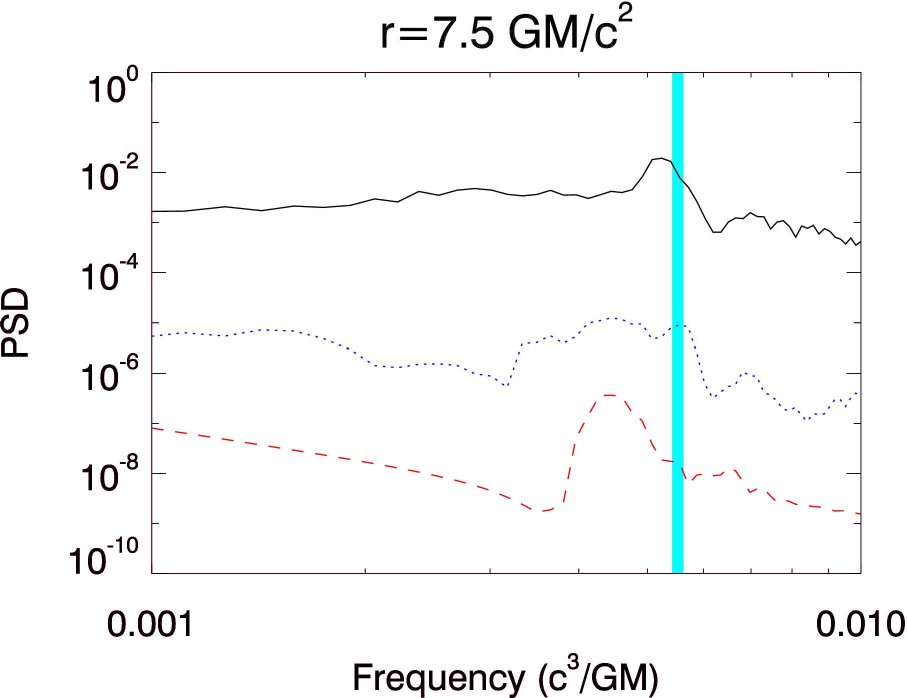}
\includegraphics[type=pdf,ext=.pdf,read=.pdf,width=0.5\textwidth]{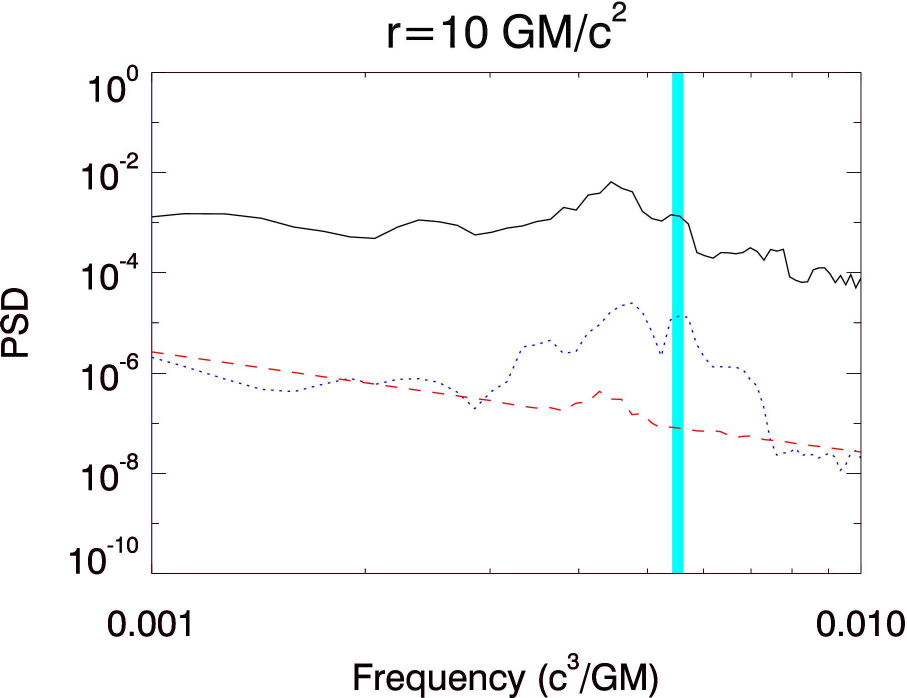}
\\[0.75em]
\includegraphics[type=pdf,ext=.pdf,read=.pdf,width=0.5\textwidth]{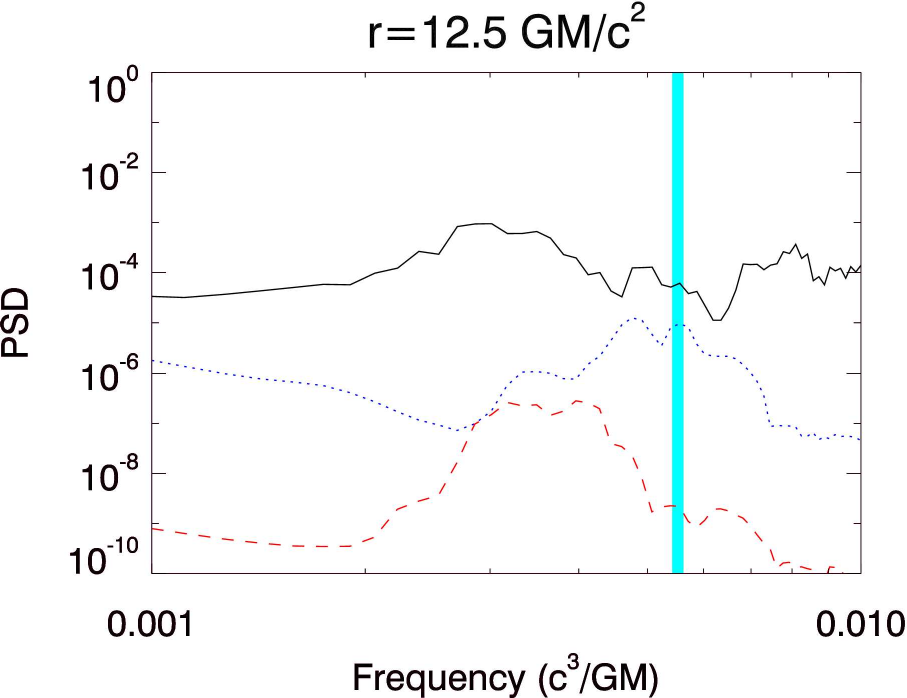}
\includegraphics[type=pdf,ext=.pdf,read=.pdf,width=0.5\textwidth]{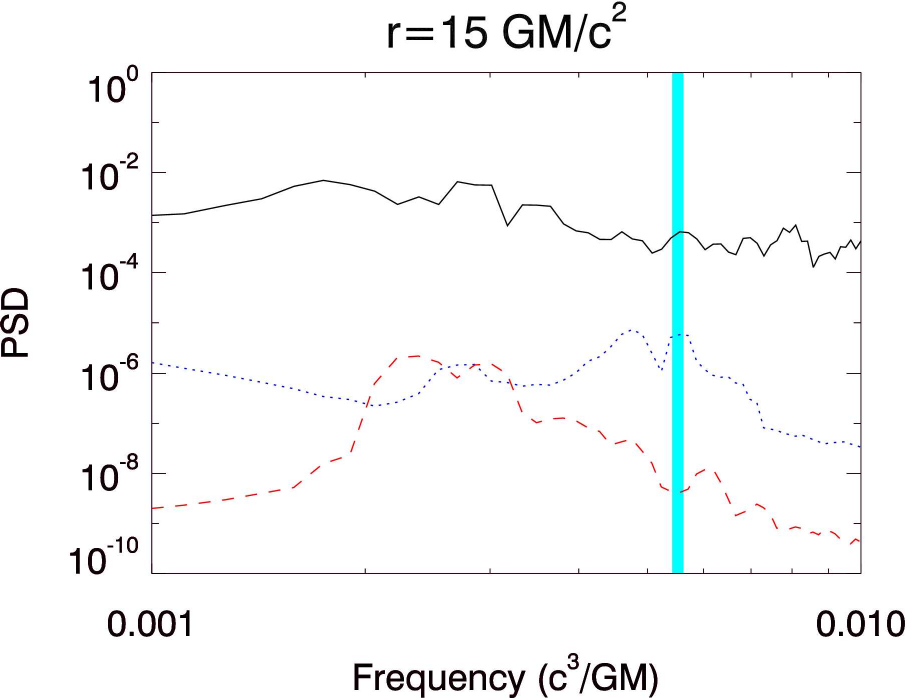}
\caption[Comparison of radial velocities]{Midplane PSDs of radial velocity, summed over radial ranges $\Delta r = 0.5r_{\rm g}$ and each centered at the listed radius.  Shown are models MHD\_{\rm 1} (solid), A0.1 (dotted), and A0.01 (dashed).  The vertical line indicates the position of the maximum radial epicyclic frequency.  In all cases, the signals of trapped modes in our model viscous disks would fall at least an order of magnitude below the noise level in MHD\_{\rm 1}.}
\label{fig:vrcomp}
\end{figure*}

Figures \ref{fig:pcomp} and \ref{fig:vrcomp} show midplane PSDs of (decay-corrected) gas pressure and radial velocity that compare MHD\_{\rm 1} (solid), A0.1 (dotted), and A0.01 (dashed).
Each plot is constructed by summing the power in that quantity over a radial range $\Delta r = 0.5 r_{\rm g}$ centered at the listed radius, and the powers are constructed so that cross-comparison between models in a given figure is valid.
In Figure \ref{fig:pcomp}, we first notice the remaining secular trend in our simulated data that scales in power roughly as $1/\omega^2$.
The imperfect process of removing this variation has left enough of this accretion-related signal present that it dominates the overall data trend, particularly in the two inner radial bins.
On top of this signal, however, we do see some oscillations that, in the case of model A0.1, are associated with trapped modes.
In the three bins centered at $r \ge 10~GM/c^3$, we see two pronounced peaks in A0.1 that correspond to the waves seen in Figures \ref{fig:a1em1psds}-\ref{fig:a1em1vrmid}, for example.
At the frequencies of interest near $\nu_{\rm max}$, however, this wave signal is always an order of magnitude or more below the noise level of model MHD\_{\rm 1}, which itself shows no convincing trapped mode signal.
Similarly, there remains no clear trapped mode signal in model A0.01.
Although Figure \ref{fig:vrcomp} features no pronounced secular trend, we again see that the trapped-mode and wave signal in model A0.1 is typically more than an order of magnitude below the MHD\_{\rm 1} noise level at frequencies near $\nu_{\rm max}$.
In this case, model A0.01 does show some oscillations near $\nu_{\rm max}$, but these are not obviously indicative of trapped mode signals and are always at least an order of magnitude below the signal in A0.1.
Taken together, these two figures suggest that a trapped mode signal corresponding to an effective viscosity of the magnitude suggested by observations \citep{2007MNRAS.376.1740K} would not be easily seen above the noise in a real MHD disk.

Our prospects of easily detecting a trapped mode signal in MHD\_{\rm 1}, however, are even further reduced because of that model's low effective viscosity.
By measuring the correlated stresses, we can estimate an effective alpha viscosity as described in \citet{1998RvMP...70....1B} and \citet{2008MNRAS.383..683P}, for example, and given by
\begin{equation}
\alpha \sim \frac{1}{\bar{p}} \left[ \langle \rho \delta v_{\rm r} \delta v_{\rm \phi} \rangle - \frac{1}{4\pi} \langle \delta B_{\rm r} \delta B_{\rm \phi} \rangle \right] .
\label{eq:alphaeff}
\end{equation}
Applying this estimator to MHD\_{\rm 1}, we find that $\alpha_{\rm MHD} \sim 0.01$, although the variation in this quantity is comparable to its value.
Still, that $\alpha_{\rm MHD} \ll 0.1$ reflects the known discrepancy between simulated and observationally inferred $\alpha$ values described in \citet{2007MNRAS.376.1740K} and further ensures that MHD\_{\rm 1} does not show a peak near $\nu_{\rm max}$.
More reflective of the inferred viscosity in MHD\_{\rm 1} is the A0.01 model, for which we have already noted the absence of any pronounced trapped mode signal.
With the addition of turbulent noise at the level present for MHD\_{\rm 1}, we can safely conclude that this model would not generate a detectable signal in a simulated MHD disk.

Although the trapped mode signal present in our 2D viscous disks could not be detected in a full 3D MHD system, we interpret this more as a shortcoming of the approach than as evidence for a dearth of diskoseismic modes in real astrophysical systems.
In addition to the basic discrepancy reported on by \citet{2007MNRAS.376.1740K}, several groups have begun to address the limitations of the current generation of MHD simulations.
\citet{2008A&A...487....1B}, for example, have noted a dependence upon computational grid resolution and, surprisingly, grid aspect ratio on values of $\alpha$ inferred from shearing-box simulations of the MRI.
Similarly, both \citet{2007ApJ...668L..51P} and \citet{2007A&A...476.1113F} have recently discussed lack of convergence in zero net magnetic flux shearing-box MHD simulations.
Specifically, they note that the effective alpha viscosities derived from these simulations decrease with increasing numerical resolution, suggesting that the saturation behavior of the MRI has yet to be captured properly.
\citet{2007ApJ...668L..51P} further point out that the physical scales of dissipation in real disks would be still smaller than the numerical resolution limit, thus making the effective viscosity in analogous real systems completely negligible.
In these cases, as in the case of MHD\_{\rm 1} and the global simulations of \citetalias{2008arXiv0805.2950R}, it is plausible that field cancellation in the absence of net magnetic flux produces an artificially low effective viscosity.
Although a strong net vertical field has the potential to disrupt the trapped g-mode region \citep{2008arXiv0806.1938F}, this process would not affect trapped p-modes.
It is also possible that one needs only a modest, and therefore non-disruptive, net vertical field to seed sufficient turbulence to produce a higher effective viscosity. 
That said, a factor of ten increase in $\alpha$ from its inferred value in MHD\_{\rm 1} would still leave diskoseismic modes at least an order of magnitude below the current turbulent MHD noise level, making them quite challenging to detect.

The presence of turbulent noise in MHD\_{\rm 1} highlights one significant way that full MHD simulations are different from alpha-disks.
Our simulated alpha-disks are characteristically non-turbulent, particularly as the viscosity increases.
Obviously, this makes the detection of diskoseismic modes in alpha-disks simpler because they feature less competing background noise than the MHD case.
This problem can in part be overcome by conducting MHD simulations over longer times to produce a better diskoseismic mode signal-to-noise ratio, although one must make certain that the integrated mass loss does not significantly change the total mass of the disk over the simulation time.
Such explorations in fact may be the only way to correctly ascertain whether the trapped modes are hidden beneath the noise or actively damped, as suggested by \citet{2006ApJ...645L..65A}.
Finally, we note that \citet{2008MNRAS.383..683P} have pointed out another shortcoming of alpha-disks, namely that real MRI-induced stresses are not typically proportional to the local shear.
All of these issues suggest that full MHD treatments are preferable when net flux simulations cease to be technically prohibitive.

\section{Conclusions}\label{sec:conclusions}
We have conducted an ensemble of axisymmetric simulations of black hole viscous accretion disks to explore the generation of diskoseismic modes and their influence on disks.
While we are still far from a definitive identification of the origin of HFQPOs, we have uncovered and explored several interesting facets of viscous disk evolution, and we summarize our findings here:

1) For viscous disks with $\alpha \geq 0.05$, we see indications of the trapped diskoseismic modes of \citet{2000ApJ...537..922O}.
These modes have all of the expected properties of trapped g-modes or inner p-modes, and are located at $r \lesssim r_{\rm max}$ with frequencies $\nu \sim \nu_{\rm max}$.
This confirms that modes similar to those seen in earlier simulations of vertically integrated models are present for 2D optically- and geometrically-thin accretion disks.

2) We note that viscous disk models with trapped diskoseismic modes also develop related waves that pervade much of the body of the disk.
These outward-propagating waves are continuous extensions in frequency and power of the trapped modes, despite extending beyond the region of formal mode trapping.
This too is similar to the 1D result and suggests that diskoseismic modes can effectively communicate their characteristic frequencies to portions of the disk in which the modes themselves would be strongly damped.

3) By comparing our viscous disks to a full 3D MHD simulation of \citetalias{2008arXiv0805.2950R}, we have further shown that the trapped mode signal for the corresponding alpha disk would fall far below the current noise level of the MHD simulation.
This suggests that, to produce detectable trapped modes, MHD simulations may need to feature larger effective viscosities, possibly through the natural incorporation of net magnetic flux.
Alternately, larger trapped mode signal-to-noise ratios should be achievable by extending the time domain of these simulations.

\acknowledgments
We wish to thank the NCSA at the University of Illinois in Urbana-Champaign for developing ZEUS-MP.  
Additionally, we wish to thank Bob Wagoner for his helpful discussions that influenced the development of this project.  
We also thank the anonymous referee for their comments.  
Some of the simulations described were run on the ``Deepthought'' High Performance Computing Cluster maintained by the Office of Information Technology at the University of Maryland, College Park.  
SMO, CSR, and MCM acknowledge the support of National Science Foundation Grant AST 06-07428.

\bibliography{refs}

\end{document}